\documentclass[fleqn,usenatbib]{mnras}

\usepackage{newtxtext,newtxmath}

\usepackage[T1]{fontenc}

\DeclareRobustCommand{\VAN}[3]{#2}
\let\VANthebibliography\thebibliography
\def\thebibliography{\DeclareRobustCommand{\VAN}[3]{##3}\VANthebibliography}

\usepackage{graphicx}	
\usepackage{amsmath}	

\usepackage{orcidlink}
\usepackage{rotating}
\usepackage{xspace}
\usepackage{ulem}
\makeatletter 
  \patchcmd{\NAT@citex}
    {\@citea\NAT@hyper@{%
      \NAT@nmfmt{\NAT@nm}%
      \hyper@natlinkbreak{\NAT@aysep\NAT@spacechar}{\@citeb\@extra@b@citeb}%
      \NAT@date}}
    {\@citea\NAT@nmfmt{\NAT@nm}%
    \NAT@aysep\NAT@spacechar\NAT@hyper@{\NAT@date}}{}{}

  \patchcmd{\NAT@citex}
    {\@citea\NAT@hyper@{%
      \NAT@nmfmt{\NAT@nm}%
      \hyper@natlinkbreak{\NAT@spacechar\NAT@@open\if*#1*\else#1\NAT@spacechar\fi}%
        {\@citeb\@extra@b@citeb}%
      \NAT@date}}
    {\@citea\NAT@nmfmt{\NAT@nm}%
    \NAT@spacechar\NAT@@open\if*#1*\else#1\NAT@spacechar\fi\NAT@hyper@{\NAT@date}}
    {}{}
\makeatother

\newcommand{\HM}{\ion{H}{$_2$}\xspace}
\newcommand{\HI}{\ion{H}{I}\xspace}
\newcommand{\HII}{\ion{H}{II}\xspace}
\newcommand{\HeI}{\ion{He}{I}\xspace}
\newcommand{\HeII}{\ion{He}{II}\xspace}
\newcommand{\HeIII}{\ion{He}{III}\xspace}

\newcommand{\thesan}{\textsc{thesan}\xspace}
\newcommand{\thzoom}{\mbox{\textsc{thesan-zoom}}\xspace}
\newcommand{\thesanone}{\mbox{\textsc{thesan-1}}\xspace}

\title[Galaxy sizes in the early Universe]{The \thzoom project: central starbursts and inside-out quenching govern galaxy sizes in the early Universe}

\author[W. McClymont et al.]{%
William McClymont $\orcidlink{0009-0009-5565-3790}$,$^{1,2}$\thanks{E-mail: \href{mailto:wjm50@cam.ac.uk}{wjm50@cam.ac.uk} (WM)}
Sandro Tacchella $\orcidlink{0000-0002-8224-4505}$,$^{1,2}$
Aaron Smith $\orcidlink{0000-0002-2838-9033}$,$^{3}$
Rahul Kannan $\orcidlink{0000-0001-6092-2187}$,$^{4}$
\newauthor
Ewald Puchwein $\orcidlink{0000-0001-8778-7587}$,$^{5}$
Josh Borrow  $\orcidlink{0000-0002-1327-1921}$,$^{6}$
Enrico Garaldi $\orcidlink{0000-0002-6021-7020}$,$^{7,8}$
Laura Keating $\orcidlink{0000-0001-5211-1958}$,$^{9}$
Mark Vogelsberger $\orcidlink{0000-0001-8593-7692}$,$^{10}$
\newauthor
Oliver Zier $\orcidlink{0000-0003-1811-8915}$,$^{11,10}$
Xuejian Shen $\orcidlink{0000-0002-6196-823X}$,$^{10}$
and Filip Popovic $\orcidlink{0009-0006-8856-918X}$$^{4}$
\\
\\
$^{1}$Kavli Institute for Cosmology, University of Cambridge, Madingley Road, Cambridge CB3 0HA, UK\\
$^{2}$Cavendish Laboratory, University of Cambridge, 19 JJ Thomson Avenue, Cambridge CB3 0HE, UK\\
$^3$ Department of Physics, The University of Texas at Dallas, Richardson, TX 75080, USA \\
$^4$ Department of Physics and Astronomy, York University, 4700 Keele Street, Toronto, ON M3J 1P3, Canada \\
$^5$ Leibniz-Institut f\"ur Astrophysik Potsdam, An der Sternwarte 16, 14482 Potsdam, Germany \\
$^6$ Department of Physics and Astronomy, University of Pennsylvania, 209 South 33rd Street, Philadelphia, PA 19104, USA \\
$^7$ Kavli Institute for the Physics and Mathematics of the Universe, The University of Tokyo, 5-1-5 Kashiwanoha, Kashiwa, 277-8583, Chiba, Japan \\
$^8$ Institute for Fundamental Physics of the Universe, via Beirut 2, 34151 Trieste, Italy \\
$^9$ Institute for Astronomy, University of Edinburgh, Blackford Hill, Edinburgh, EH9 3HJ, UK \\
$^{10}$ Department of Physics, Kavli Institute for Astrophysics and Space Research, Massachusetts Institute of Technology, Cambridge, MA 02139, USA \\
$^{11}$ Center for Astrophysics $|$ Harvard $\&$ Smithsonian, 60 Garden Street, Cambridge, MA 02138, USA
}

\date{Accepted XXX. Received YYY; in original form ZZZ}

\pubyear{\the\year{}}

\begin{document}
\label{firstpage}
\pagerange{\pageref{firstpage}--\pageref{lastpage}}
\maketitle

\begin{abstract}
We explore the evolution of galaxy sizes at high redshift ($3<z<13$) using the high-resolution \thzoom radiation-hydrodynamics simulations, focusing on the mass range of $10^6\,\mathrm{M}_{\odot} < \mathrm{M}_{\ast} < 10^{10}\,\mathrm{M}_{\odot}$. Our analysis reveals that galaxy size growth is tightly coupled to bursty star formation. Galaxies above the star-forming main sequence tend to form stars in a central starburst, which decreases their radial size. These galaxies quench inside-out, causing spatially extended star formation and increasing their radial size, leading to oscillatory behavior around the size--mass relation. Notably, we find a positive intrinsic size--mass relation at high redshift, consistent with observations but in tension with large-volume simulations. We attribute this discrepancy to the bursty star formation captured by our multi-phase interstellar medium framework, but missing from simulations using the effective equation-of-state approach with hydrodynamically decoupled feedback. We also find that the normalization of the size--mass relation follows a double power law as a function of redshift, with a break at $z\approx6$, because the majority of galaxies at $z>6$ show rising star-formation histories, and therefore are in a compaction phase. We demonstrate that H$\alpha$ emission is systematically extended relative to the UV continuum by a median factor of 1.7, consistent with recent \textit{JWST} studies. However, in contrast to previous interpretations that link extended H$\alpha$ sizes to inside-out growth, we find that Lyman-continuum (LyC) emission is spatially disconnected from H$\alpha$. Instead, a simple Str\"{o}mgren sphere argument reproduces observed trends, suggesting that extreme LyC production during central starbursts is the primary driver of extended nebular emission.
\end{abstract}




\section{Introduction}
\label{sec:Introduction}

As galaxies assemble and build up their mass, so too do they evolve in physical size. Empirically, the relationship between stellar mass and size evolution is characterized by the size--mass scaling relation, which follows a simple power-law $R \propto M_\ast^\alpha$ \citep{Kauffmann:2003aa,Shen:2003aa}. However, the large intrinsic scatter in this relation indicates that galaxy sizes do not simply trace mass build-up, but are also sensitive to a wide array of complex processes governing galaxy evolution, including feedback-regulated star formation and mergers. This is most easily seen in the clearly distinct size--mass relations of star-forming galaxies and quiescent galaxies, implying that size growth is strongly tied to a galaxy's star-formation history \citep[SFH;][]{Franx:2008aa,Lilly:2016aa,Nedkova:2021aa}. 

While it is natural to describe galaxy sizes in terms of their mass distribution, through quantities such as the stellar half-mass radius, observationally we are only able to directly probe light and therefore galaxy sizes are usually defined across a range of wavelengths. This introduces a number of complications in understanding the mass distribution of galaxies, as the mass-to-light ratio varies drastically depending on the properties of the stellar population and on the particular wavelength range observed \citep{Wuyts:2010aa,Szomoru:2013aa,Lang:2014aa,Mosleh:2017aa,Suess:2019aa,Papaderos:2023aa}. However, this can also prove advantageous, as probing galaxy sizes with a variety of wavelengths allows us to understand the spatial distribution of different stellar populations \citep{Wuyts:2013aa,Morselli:2019aa,Jain:2024aa}. For example, UV light is generally dominated by young stars, whereas stars can remain bright in the optical over a much longer period of time. Comparing UV and optical sizes therefore allows us to understand whether recent star formation has occurred on a larger or smaller scale than the older stellar population. Caution is needed, however, as dust acts to redden the observed light, and so spatially varying dust attenuation will also have a wavelength-dependent effect on observed sizes \citep{Nelson:2016aa,Tacchella:2018ab}. 

The study of wavelength-dependent sizes can be extended even further by isolating wavelengths associated with bright nebular line emission, allowing us to study the morphology of ionized gas \citep{Nelson:2013aa,Nelson:2016ab, Tacchella:2015aa,Forster-Schreiber:2018aa}. In typical star-forming galaxies, the dominant source of dense ionized gas is photoionization by the copious Lyman continuum (LyC) emission from young, massive stars. We can therefore use the morphology of nebular emission to trace ongoing star formation, provided that we assume LyC photons are absorbed locally. The validity of this assumption is complicated by the large fraction of nebular emission arising from non-local diffuse ionized gas extending beyond active star-forming regions, although the effect on nebular sizes is likely small for massive star-forming galaxies in the local Universe which have large sizes relative to the mean-free path of ionizing photons \citep{Belfiore:2022aa,Smith:2022aa,Tacchella:2022aa,McClymont:2024aa}.

The size--mass relation is expected to evolve with redshift due to a wide variety of processes including mergers \citep{Naab:2007aa,Buitrago:2008aa,Naab:2009aa,Newman:2012aa,Oser:2012aa,McIntosh:2014aa} and feedback \citep{Fan:2008aa,Damjanov:2009aa,Tacchella:2016ab}. Studies of the UV size--mass relation out to $z\approx8$ with the \textit{Hubble Space Telescope (HST)} have found a rapid evolution of the normalization with redshift \citep{Oesch:2010aa,Bruce:2012aa,Mosleh:2012aa,Ono:2013aa,Morishita:2014aa,van-der-Wel:2014aa,Shibuya:2015aa,Allen:2017aa,Mosleh:2020aa,Bouwens:2022aa}. In recent years, the study of galaxy morphology has been revolutionized by the \textit{James Webb Space Telescope (JWST)}, which has an unprecedented sensitivity, angular resolution, and wavelength coverage. This is facilitating detailed morphological studies \citep{Baker:2024aa}, large sample analysis of galaxy morphology during the Epoch of Reionization \citep{Kartaltepe:2023aa,Ormerod:2024aa,Morishita:2024aa,Sun:2024aa,Miller:2025aa,Allen:2025aa,Yang:2025ab}, and the measurement of galaxy sizes out to the new redshift frontier at $z\approx14$ \citep{Robertson:2024aa}. \textit{JWST} has revealed a large population of compact galaxies at high redshift, and initial studies have generally found a strong evolution in the size--mass relation and spatially consistent optical and UV emission \citep{Sun:2024aa,Ormerod:2024aa,Miller:2025aa}.

A great deal of work has been carried out investigating the physical mechanisms which drive galaxy sizes and morphology using hydrodynamic simulations, and has been remarkably successful at reproducing the observed features of galaxies in the local Universe \citep{Furlong:2017aa,Genel:2018aa}. However, simulations have generally struggled to reproduce the more extreme compact galaxies observed at high-redshift and their size--mass relations are in tension with \textit{JWST} results \citep{Shen:2024aa}. Additionally, large-volume simulations tend to predict a negative slope in the size--mass relation at high redshift, which is at odds with observed size--mass relations \citep{Marshall:2022aa,Roper:2022aa,Shen:2024aa}, although it has been argued that centrally concentrated dust in massive galaxies may be able to reconcile predicted and observed size--mass relations by preferentially attenuating light near the galactic center \citep{Wu:2020aa,Marshall:2022aa,Popping:2022aa,Roper:2022aa,Costantin:2023aa}.

\textit{JWST} has also ushered in a revolution in multi-wavelength size analysis in the early Universe, greatly expanding the redshift horizon for optical size measurements and allowing for nebular galaxy sizes to be studied through the use of medium band photometry (Villanueva et al. in prep.) and through the NIRCam Grism (Danhaive et al. in prep.). These works have generally found extended nebular sizes relative to the continuum, in agreement with other works that find increasing nebular line equivalent widths (EWs) with radius at high redshift \citep{Nelson:2024aa,Matharu:2024aa} and in local starbursts \citep{Papaderos:2002aa,Papaderos:2012aa}. Extended nebular sizes and radially increasing EWs have generally been interpreted as indicators of inside-out growth in high-redshift galaxies, where the core of a galaxy is formed first and subsequent star formation occurs at successively larger radii \citep{Nelson:2024aa,Matharu:2024aa}, however in this work we will conclude that extended nebular emission at high redshift is instead caused by radiative transfer effects related to the bursty nature of star formation at these high redshifts. Interestingly, \textit{JWST} has also revealed a population of anomalous Balmer emitters (ABEs), which are galaxies exhibiting line ratios inconsistent with Case B recombination \citep{Scarlata:2024aa,Yanagisawa:2024aa,McClymont:2025ad}. One interpretation of ABEs is that they are density-bounded nebulae caused by an intense central starburst \citep{McClymont:2025ad}, which we should expect to show extended nebular sizes relative to the stellar continuum if the ionizing photons are being absorbed by gas outside the interstellar medium (ISM). Understanding the link between bursty star formation and nebular sizes is particularly useful in the context of recent studies of bursty star formation through both observations \citep{Looser:2024aa,Looser:2025aa,Endsley:2024ab,Endsley:2025aa,Baker:2025aa,Trussler:2025aa,Witten:2025aa} and theory \citep{Hayward:2017aa,Tacchella:2020aa,Hopkins:2023aa,McClymont:2025aa} as it may provide insight into where star formation is concentrated throughout the burst cycle. 

Concurrent with new observational measurements of nebular galaxy sizes, the generation of nebular line emission from simulations has advanced significantly in recent years. Mock images of line emission from simulated galaxies can now be sourced directly from gas cells thanks to both on-the-fly and post-processed radiative transfer techniques \citep[e.g.][]{Smith:2022aa,Katz:2022ad}. This is a drastic improvement for measuring sizes over the more common method of tying nebular emission to young stellar particles with a precalculated photoionization model \citep{Hirschmann:2017aa,Shen:2020aa,Kannan:2022ab,Yang:2023aa,Yang:2025aa}. This development allows for nebular emission which is spatially separated from young stars to be self-consistently accounted for, such as emission from the diffuse ionized gas and ionized outflows, and therefore allows for more accurate nebular size measurements in simulations \citep{Tacchella:2022aa,McClymont:2024aa,McCallum:2024aa}.

In this work, we use the high-resolution \thzoom simulations \citep{Kannan:2025aa} to investigate the evolution of the galaxy size--mass relation at $3<z<13$. We show that our simulated galaxies agree well with recent \textit{JWST} measurements. We explore the apparent and actual size evolution of galaxies with redshift and investigate how galaxies morphologically grow in the earliest stages of galaxy evolution, with particular reference to bursty star formation. We also investigate the relationship between H$\alpha$ sizes and continuum sizes in detail, showing that extended H$\alpha$ sizes are closely related to the bursty nature of intense star formation in the early Universe.

In Section~\ref{sec:Simulation methodology} we describe the \thzoom simulations, the selection of our galaxy sample, and and our post-processing methods. In Section~\ref{sec:Galaxy sizes} we show the \thzoom size--mass relations and compare them to the literature. In Section~\ref{sec:The size--mass relation across cosmic time} we discuss the redshift evolution of galaxy sizes. In Section~\ref{sec:Extended nebular emission} we explore the processes that govern the nebular sizes of galaxies. In Section~\ref{sec:Summary and Conclusions} we summarize our main conclusions.

\section{Simulation methodology}
\label{sec:Simulation methodology}

\begin{figure*} 
\centering
	\includegraphics[width=\textwidth]{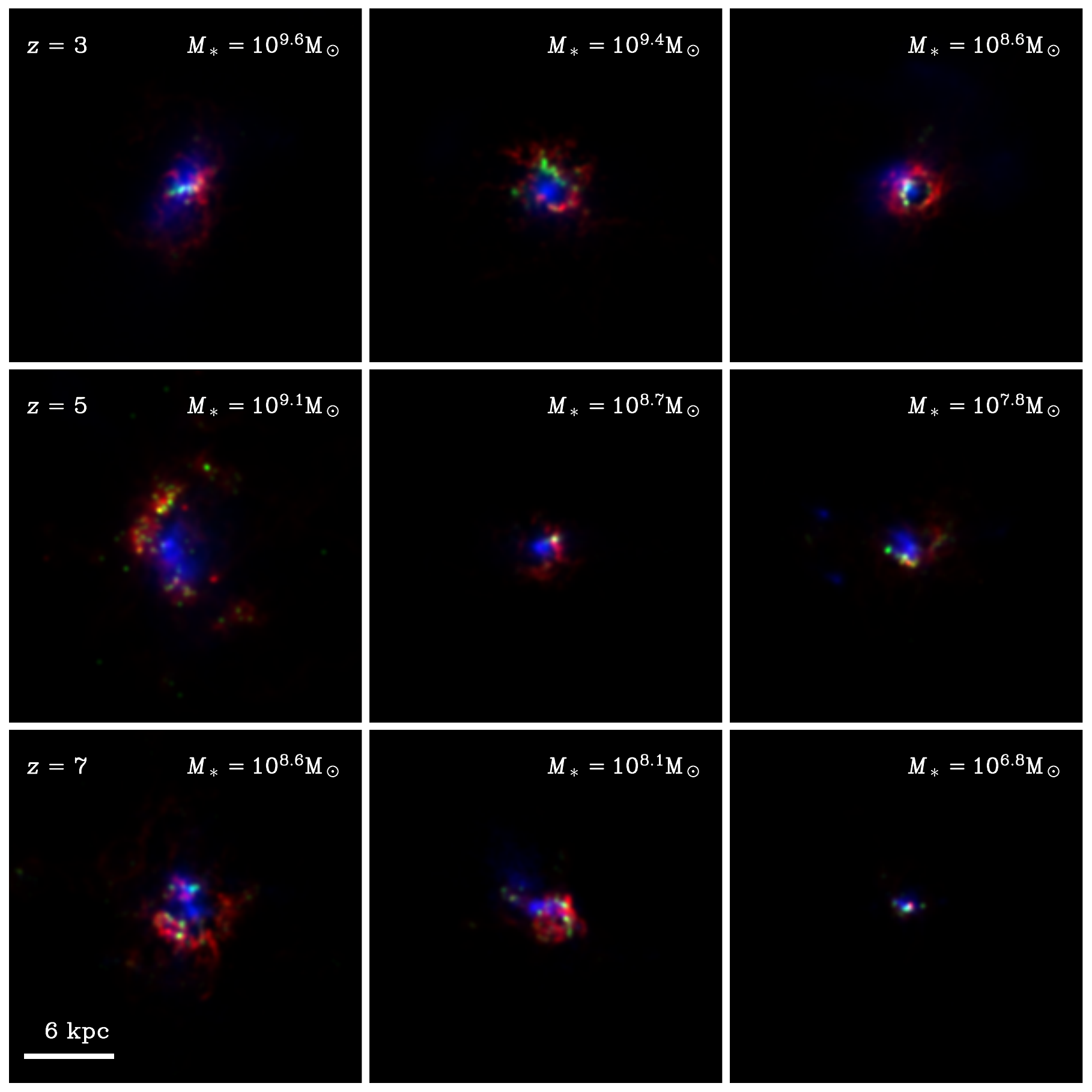}
    \caption{RGB images of 3 \thzoom galaxies at three redshifts in intrinsic H$\alpha$ (red), Lyman continuum (LyC, green), and UV (blue). We show m12.6 (left, subhalo 3), m12.6 (middle, subhalo 4), and 11.1 (right, subhalo 0). The different tracers generally show remarkably distinct morphologies. H$\alpha$ traces ionized gas and therefore appears near LyC clumps, however a strong H$\alpha$ haze can also be identified around several galaxies. LyC emission tends to arise in compact star-forming regions scattered around the galaxy. The UV morphology is smoother than LyC, owing to the fact that stars from previous bursts can still contribute strongly to UV emission.}
    \label{fig:thumbnail_images_6x6}
\end{figure*}

\subsection{Simulations and galaxy sample}
\label{sec:Simulations and galaxy sample}

This work employs the \thzoom simulations, a cutting-edge suite of high-resolution zoom-in simulations designed to address the limitations of large-scale cosmological projects like \thesan \citep{Kannan:2022aa, Smith:2022ab, Garaldi:2022aa}. While \thesan modeled galaxy formation and the Epoch of Reionization (EoR) over $\sim$100\,cMpc volumes using the IllustrisTNG framework \citep{Pillepich:2018aa} with on-the-fly radiative transfer, their effective equation-of-state approach to star-forming gas restricts detailed exploration of processes within galaxies. \thzoom addresses this gap by focusing on smaller regions with enhanced resolution and a more sophisticated ISM model and stellar feedback to explore and improve predictions affected by such modeling \citep[e.g.][]{Yeh:2023aa,Garaldi:2024aa,Shen:2024aa}

The \thzoom simulations use a ``zoom-in'' technique to select target regions from the \thesan parent volume, allowing detailed investigation of galaxy-scale phenomena within the broader cosmological context. A key feature is the incorporation of the time-varying radiation field from the original \thesan simulation as a boundary condition, enabling realistic interactions between galaxies and their external radiation environments. This feature is crucial for studying galaxies at high redshift, where radiative feedback from neighboring sources is thought to play a significant role in shaping their evolution \citep{Rosdahl:2018aa, Ocvirk:2020aa}.

Full technical details of the simulations are presented in \citet{Kannan:2025aa}, though we provide a summary here. The simulations have three resolution levels which correspond to factors of  4, 8, and 16 improvements in the spatial resolution and factors of 64, 512, and 4096 in the mass resolution relative to \thesan. The baryonic mass resolutions are $9.09\times10^3\,\mathrm{M}_\odot$, $1.14\times10^3\,\mathrm{M}_\odot$, and $1.42\times10^2\,\mathrm{M}_\odot$ respectively. In this work we always use the highest resolution fiducial run for each zoom region.

\thzoom utilizes the {\sc arepo-rt} radiation hydrodynamics code \citep{Kannan:2019aa}, built on the moving-mesh framework {\sc arepo} \citep{Springel:2010aa}, and uses a novel node-to-node communication strategy \citep{Zier:2024aa}. Radiative transfer is computed on the fly using a moment-based scheme that tracks the photon number density and flux while employing a reduced speed of light approximation to improve computational efficiency. The simulations model non-equilibrium thermochemical processes for six species ($\HM, \HI, \HII, \HeI, \HeII,$ and $\HeIII$). Metal cooling rates, precomputed using \textsc{cloudy} \citep{Ferland:2017aa}, assuming ionization equilibrium with a \cite{Faucher-Giguere:2009aa} UV background and are stored in look-up tables \citep{Vogelsberger:2013aa}.

A significant strength of \thzoom is its advanced treatment of stellar feedback, which combines multiple mechanisms including photoionization, radiation pressure, stellar winds, and supernova explosions \citep{Kannan:2020aa, Kannan:2021aa}. Young stars contribute ionizing radiation that directly heats and ionizes the surrounding gas. This radiative feedback is self-consistently implemented through the on-the-fly radiation transfer. Stellar winds are implemented with the SMUGGLE model as described in \citep{Marinacci:2019aa}. Supernova feedback injects momentum and thermal energy into the ISM. To address the overproduction of stars in dense regions, an early stellar feedback mechanism disrupts molecular clouds shortly after star formation begins, effectively regulating star formation rates. These feedback processes collectively shape the properties of nascent galaxies. There have been a number of initial works using \thzoom simulations, including galaxy-scale star-formation efficiencies \citep[SFEs;][]{Shen:2025aa}, the relationship between reionization and galaxy properties \citep{Zier:2025aa}, Population III stars \citep{Zier:2025ab}, galaxy sizes \citep{McClymont:2025ab}, giant molecular clouds (GMC) scale SFEs \citep{Wang:2025aa}, and galactic chemical evolution \citep{McClymont:2025ae}.

Stellar masses and SFRs are calculated based on the bound particles within the virial radius for each subhalo. SFRs are calculated based on recently formed stars within the relevant averaging timescale ($t_\mathrm{avg}$) as
\begin{equation}
  \mathrm{SFR}_{t_\mathrm{avg}} = \frac{\sum_im_{\ast,i}}{t_\mathrm{avg}}\,,
\end{equation}
where we are summing over each stellar particle with an age less than $t_\mathrm{avg}$, and where $m_{\ast,i}$ is the initial mass of each particle. We define $t_\mathrm{avg}$ in Myr.

There are 14 zoom regions, centered on target haloes chosen to span a wide range of $z=3$ halo masses ($M_\mathrm{halo}\approx10^8-10^{13}\,\mathrm{M_\odot}$). In this work, we consider subhaloes which are resolved with at least 100 stellar particles. The friends-of-friends (FOF) algorithm was used to identify haloes \citep{Davis:1985aa}, with self-gravitating subhalos being identified within the FOF groups using the SUBFIND-HBT algorithm \citep{Springel:2001aa,Springel:2021aa}. We include both central and satellites in this analysis and include all galaxies with $10^5\,\mathrm{M}_\odot<M_\ast<10^{11}\,\mathrm{M}_\odot$ at $3<z<13$, leaving us with a sample of 130762 subhaloes, comprised of 1781 unique trees. For the individual snapshot at $z=3$ there are 551 subhaloes and at $z=13$ there are 59 subhaloes. We follow the naming scheme for zoom regions presented in \citet{Kannan:2025aa}.

\subsection{Radiative transfer of non-ionizing continuum and emission line photons}
\label{sec:Radiative transfer of non-ionizing continuum and emission line photons}

In this study, we focus on the H$\alpha$, optical continuum, and UV continuum emission from galaxies. To calculate both continuum and line luminosities, we utilize the Monte Carlo radiative transfer (MCRT) Cosmic Ly$\alpha$ Transfer code \citep[\textsc{colt};][]{Smith:2015aa, Smith:2019aa, Smith:2022aa}. Fig.~\ref{fig:thumbnail_images_6x6} presents example RGB images for the \thzoom galaxies in H$\alpha$, LyC, and UV emission. We note that where LyC emission is considered in this work, we are exclusively interested in the intrinsic emission and therefore do not consider the radiative transfer of ionizing radiation, which is presented in upcoming studies. The ionization states are taken directly from gas cells as calculated by the on-the-fly RT solver. Our methodology for calculating non-ionizing continuum and line emission largely follows that outlined in previous works \citep[e.g.][]{Smith:2022aa,Tacchella:2022aa,McClymont:2024aa}. We briefly summarize the key aspects of our implementation.

H$\alpha$ is a recombination line, and so the luminosity of a given gas cell is given by
\begin{equation}
  L_\mathrm{H\alpha}^\mathrm{rec} = h \nu_\mathrm{H\alpha} \int P_{\mathrm{B},\mathrm{H\alpha}}(T,n_e) \alpha_\mathrm{B}(T)\,n_e n_\HII\,\text{d}V \, ,
\end{equation}
the energy at line center is $h\nu_\mathrm{H\alpha}$, and $n_e$ and $n_\text{\HII}$ are the number densities for free electrons and ions, respectively. $\alpha_\mathrm{B}$ is the Case B recombination coefficient, and $P_{\mathrm{B},\mathrm{H\alpha}}$ is the probability for the emission line photon to be emitted per recombination event. The coefficient $\alpha_\mathrm{B}$ is from \citet{Hui:1997aa}. $P_{\mathrm{B},\mathrm{H\alpha}}$ is calculated from \citet{Storey:1995aa}. To prevent nonphysical Balmer line ratios that might result from unrealistically low temperatures in the ionized gas, a temperature floor of 7000\,K is applied to the calculation of $P_{\mathrm{B},\mathrm{H\alpha}}$, though we stress that this floor is not applied to the recombination rate.

The continuum emission is derived from stellar spectral energy distributions (SEDs) computed using the Binary Population and Spectral Synthesis (BPASS) model with binaries, following a Chabrier initial mass function \citep[IMF;][]{Chabrier:2003aa}, with a maximum stellar mass of $100\,\mathrm{M_\odot}$ \citep[v2.2.1;][]{Eldridge:2009aa, Eldridge:2017aa}\footnote{Further information on BPASS can be found on the project website at \href{https://bpass.auckland.ac.nz}{\texttt{bpass.auckland.ac.nz}}.}. In our analysis, we concentrate on the far-ultraviolet (FUV) and optical so we use emission in windows of 1475--1525\AA\ and 5000--6000\AA\ respectively. The intrinsic Lyman continuum (LyC) emission is similarly determined from these BPASS SEDs.

Photon packets are sampled according to the source luminosity and are assigned corresponding weights. For continuum emission, photons are launched isotropically from stellar positions, whereas for line emission, the photons are emitted isotropically from randomly selected locations within the gas cell. As these photons propagate, they encounter absorption and scattering by dust, modeled here using the fiducial two-component dust prescription described in Garaldi et al. (in prep.). In practice, we only make reference throughout this work to the intrinsic, unattenuated emission from galaxies. This is because the fiducial dust model has a completely negligible impact on galaxy sizes in the mass and redshift range considered here. For example, the ratio of observed to intrinsic UV size is $1.000^{+0.001}_{-0.004}$, where the errors are the $16^\mathrm{th}$--$84^\mathrm{th}$ percentile scatter.

\section{Galaxy sizes}
\label{sec:Galaxy sizes}

\begin{figure*} 
\centering
	\includegraphics[width=\textwidth]{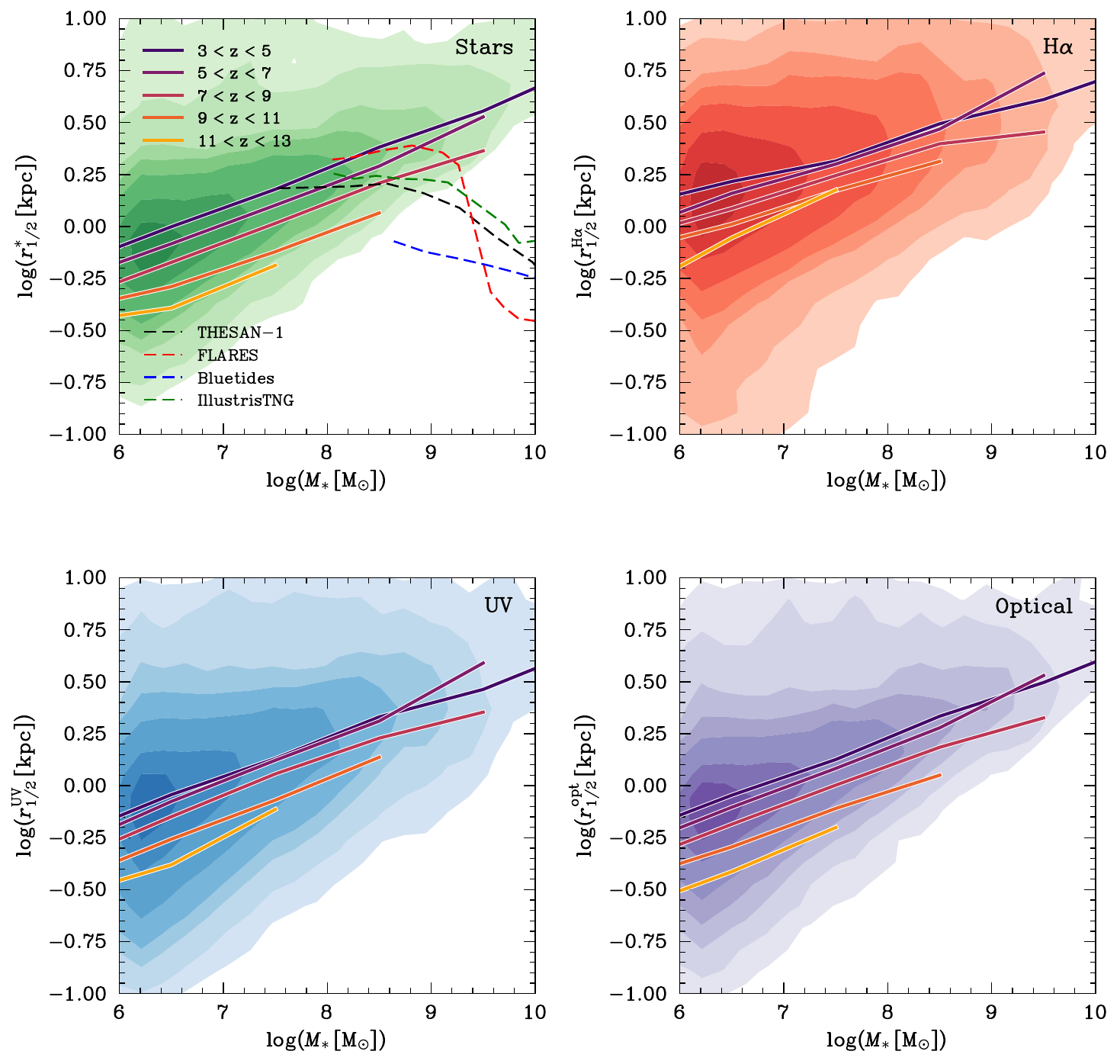}
    \caption{The half-light and stellar half-mass radii of \thzoom galaxies as a function of stellar mass. We show stellar mass ($r^\ast_{1/2}$, upper left), H$\alpha$ ($r^{\mathrm{H}\alpha}_{1/2}$, upper right), UV ($r^\mathrm{UV}_{1/2}$, lower left), and optical ($r^\mathrm{opt}_{1/2}$, lower right) sizes. The contour plots show the distribution of galaxies across $3<z<13$, with contours encompassing 20, 50, 80, 90, 95, 99, and 99.9$\%$ of the distribution. Colored lines represent the median relation in redshift bins, where at least 10 galaxies are required for each point to be plotted. We see a positive size--mass relation for the stellar half-mass across the entire stellar mass and redshift range. This naturally leads to positive size--mass relations for $r^{\mathrm{H}\alpha}$, $r^\mathrm{UV}_{1/2}$, and $r^\mathrm{opt}_{1/2}$. This is in contrast to the negative size--mass relations from \thesanone \citep{Shen:2024aa}, IllustrisTNG \citep{Genel:2018aa}, FLARES \citep{Roper:2022aa}, and Bluetides \citep{Marshall:2022aa}, where concentrated dust must be invoked to recover a positive observed size--mass relation. We show the median $r^\ast_{1/2}$ for these simulations at $5\leq z\leq7$ in the upper left panel. }
    \label{fig:stellar_mass_size}
\end{figure*}

\subsection{Measuring galaxy sizes}
\label{sec:Measuring galaxy sizes}

\begin{figure*} 
\centering
	\includegraphics[width=\textwidth]{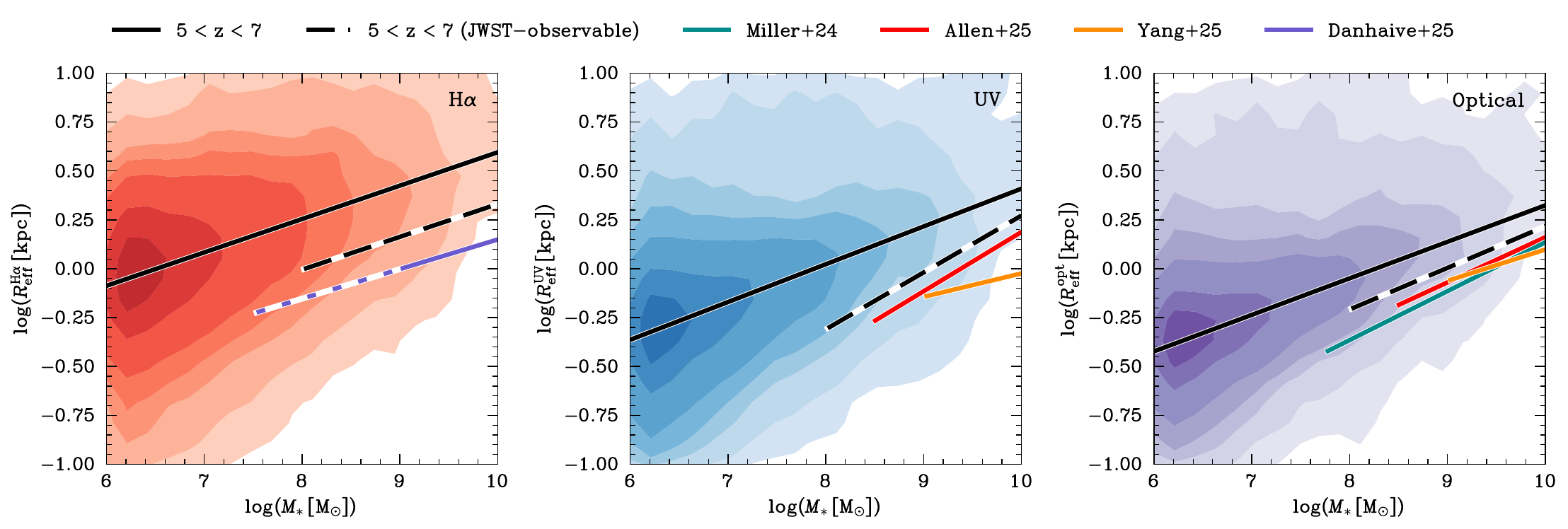}
    \caption{The intrinsic 2D half-light radii of \thesan galaxies in H$\alpha$ (left, $R^{\mathrm{H\alpha}}_{\mathrm{eff}}$), UV (center, $R^{\mathrm{UV}}_{\mathrm{eff}}$), and optical (right, $R^{\mathrm{opt}}_{\mathrm{eff}}$) emission as a function of stellar mass. The contour plots show the distribution of galaxies across $5<z<7$, with contours encompassing 20, 50, 80, 90, 95, 99, and 99.9$\%$ of the distribution. We also show our best fit line in the redshift bin $5<z<7$ (solid black) and our best fit after biasing for galaxies observable in typical \textit{JWST} surveys (dash black, see text for details). Projection effects cause 2D half-light sizes to be smaller and observations are unable to detect dim, extended galaxies. Accounting for these brings our biased best fit line within 0.1--0.2 dex of \textit{JWST} observational results for H$\alpha$ (Danhaive et al. in prep.), optical \citep{Miller:2025aa,Allen:2025aa}, and UV sizes \citep{Allen:2025aa,Yang:2025ab}. }
    \label{fig:2d_size_mass}
\end{figure*}

In order to measure the 3D half-mass or half-light centers, we construct radial mass or light-weighted histograms. We adopt the UV center–of–light as the common definition of the galactic center for all light-weighted size measurements. For the mass-weighted sizes, we instead use the subhalo center (based on the most bound particle). We define 40 log-spaced radial bins from $10^{-3}\,R_{\mathrm{vir}}$ to $10\,R_{\mathrm{vir}}$ and add a 41st bin from $0$ to $10^{-3}\,R_{\mathrm{vir}}$. In each radial bin we calculate the luminosity or mass of that bin, cumulatively add the bins, and then interpolate to find the radius corresponding to half of the total luminosity or mass. In addition to stellar half-mass sizes, we also consider SFR-weighted, H$\alpha$, optical (5000--6000~\AA), UV (1475--1525~\AA), and LyC (<13.6\,eV) sizes.

In Fig.~\ref{fig:stellar_mass_size} (upper left panel) we show the stellar half-mass radius, $r^\ast_{1/2}$, as a function of stellar mass. We also show median trends from various other simulations at $z\approx6$, including the original \thesanone simulation. The difference in these size--mass relations is striking. In the \thzoom simulations, the size--mass relation is positive at all sizes and masses considered, whereas in \thesanone \citep{Shen:2024aa}, IllustrisTNG \citep{Genel:2018aa}, FLARES \citep{Roper:2022aa}, and Bluetides \citep{Marshall:2022aa} the size--mass relation is negative at high-redshift. These simulations rely on concentrated dust attenuation to cause their observed size--mass relations to better match the positive slope, although it is unclear how well this matches observational results showing little systemic differences between optical and UV sizes at high redshift \citep{Miller:2025aa,Allen:2025aa}. We investigate this further in Section~\ref{sec:Evolution of individual galaxies}.

In Fig.~\ref{fig:stellar_mass_size} we also show the intrinsic 3D half-light sizes of galaxies as a function of stellar mass for $r^{\mathrm{H\alpha}}_{\mathrm{1/2}}$ (upper right panel), $r^{\mathrm{UV}}_{\mathrm{1/2}}$ (lower left panel), and $r^{\mathrm{opt}}_{\mathrm{1/2}}$ (lower right panel). Our galaxies are generally more extended in H$\alpha$ compared to UV, in agreement with \textit{JWST} results (Villanueva et al. in prep., Danhaive et al. in prep.). We explore this in detail in Section~\ref{sec:Extended nebular emission}.

The trends of the size--mass relations are mostly the same for $r^{\mathrm{UV}}_{\mathrm{1/2}}$, $r^{\mathrm{opt}}_{\mathrm{1/2}}$, and $r^\ast_{1/2}$, although the apparent redshift evolution of $r^{\mathrm{UV}}_{\mathrm{1/2}}$ is marginally weaker at low redshift. Such agreement between $r^{\mathrm{UV}}_{\mathrm{1/2}}$ and $r^{\mathrm{opt}}_{\mathrm{1/2}}$ has been seen previously in both observations \citep{Miller:2025aa,Allen:2025aa} and simulations \citep{Wu:2020aa}. This agreement is due to young stars dominating both the optical and UV continua in high-redshift galaxies.

We note that the selection of a galactic center can introduce systematic errors in size measurements. We use UV center-of-light for light-weighted sizes and the subhalo center for mass-weighted sizes, however, our results are encouragingly robust to reasonable choices of center. For example, if we recompute our results for light-weighted sizes using the subhalo center, we recover the same trends and normalization for the medians with mass and redshift. Additionally, the general consistency between optical and UV sizes gives us confidence that the UV center of light is an appropriate choice for the galactic center for our sample. However, it is important to note that the UV center of light is arguably least robust for galaxies which have had very little star formation for a long period, as the formation of a single stellar particle can dramatically shift the UV center of light. Across our whole sample, 12\% of galaxies have SFR$_{50}=0\,\mathrm{M_\odot yr^{-1}}$ and therefore we expect this to be a subdominant effect overall, although it certainly does lead to an extended distribution of galaxies out to very large sizes ($\sim10$\,kpc), which is visible in the contour plots. To guard against any biases this may introduce, we have checked that our use of medians and student-t likelihood fitting robustly identifies the peak of relations throughout.

\subsection{Comparison to observations}
\label{sec:Comparison to observations}

To compare our size--mass relationship with observational results, we must compute the projected 2D sizes of our galaxies. To do this, we first calculate the luminosity density, $j(r)$, in thin radial bins from the galactic center. We then perform a numerical integration to calculate the projected intensity, $I(R)$,
\begin{equation}
I(R)=2\int^{\infty}_R\frac{j(r)\,r\,dr}{\sqrt{r^2-R^2}}\,,
\end{equation}
where $r$ is the 3D radius and $R$ is the projected 2D radius. We then calculate the 2D half-light radius from this intensity function.

This procedure implicitly assumes that the emission is spherically symmetric, which is of course not valid across all galaxies. Ideally, one would generate synthetic images from multiple viewing angles, convolve with a point–spread function, add realistic noise, and fit a Sérsic profile. Our procedure does not consider these extra sources of bias and scatter. However, we consider it sufficient for this work as it captures the dominant effect of concentrating emission in the core. We also note that we are measuring half-light sizes, whereas observations often report the semi-major axis.

The projected 2D sizes of all galaxies are smaller than their 3D sizes, and the extent to which each galaxy shrinks depends on the luminosity density concentration. The median shrinking factor is $R^{\mathrm{UV}}_{\mathrm{eff}}/r^{\mathrm{UV}}_{\mathrm{1/2}}=0.66^{+0.14}_{-0.16}$, where the errors are the $16^\mathrm{th}$--$84^\mathrm{th}$ percentile scatter. Half-light radii are often used to characterize the size of galaxies for the purpose of calculating densities. The underestimation of half-light radii therefore implies that calculated volume densities are on average overestimated by a factor of 3.5.

For comparison to observations, we fit the size--mass relation following
\begin{equation}
\log(r\,\mathrm{[kpc]}) = \alpha\log\left(\frac{M_\ast}{10^8\,\mathrm{M}_\odot}\right) + \gamma\,,
\end{equation}
and restrict our fits to galaxies with $10^6\,\mathrm{M}_\odot<M_\ast<10^{10}\,\mathrm{M}_\odot$. We divide the fits into five equally spaced redshift bins between $z=3$ and $z=13$. This better mimics the approach taken by observers than our fiducial redshift-dependent fits (see Section~\ref{sec:Measuring galaxy sizes} and Tab.~\ref{tab:size_mass_fits}).

In Fig.~\ref{fig:2d_size_mass} we show the size--mass relationship for $R^{\mathrm{UV}}_{\mathrm{eff}}$, $R^{\mathrm{opt}}_{\mathrm{eff}}$, and $R^{\mathrm{H\alpha}}_{\mathrm{eff}}$. The 2D size-mass relations lie closer to observationally measured size--mass relations, however, there is still an offset of $\sim$0.3\,dex. To more fairly compare our relation with the observed relations, we must account for the observability of our galaxies. Given the range of survey depths across filters and differing selection criteria between studies \citep{Miller:2025aa,Allen:2025aa}, it is impossible to construct an ideal and universal mock observational selection unless one fully forward models all parts of the analysis pipeline. While such a forward modeling approach would be valuable, it is outside the scope of this work. Therefore, we aim instead to create a selection of galaxies which could reasonably be detected and measured with \textit{JWST}. To do this we broadly follow the procedure of \citet{Shen:2024aa}, where the quoted point-source 5$\sigma$ limiting magnitude is divided by the assumed aperture size. While \citet{Shen:2024aa} only require that the galaxy peaks above the 1$\sigma$ limit, we instead try to more closely match the much harsher selection criteria used in \citet{Allen:2025aa} and \citet{Miller:2025aa}. We assume a 5$\sigma$ survey depth of 29\,mag within an aperture size of 0.16\,arsec, giving a 5$\sigma$ surface brightness limit of 26.3\,mag\,arcsec$^{-2}$. We require a 5$\sigma$ average surface brightness within the half-light radius for both optical and UV emission. We also require that a galaxy is brighter than 29\,mag, which is necessary to ensure detectability for galaxies smaller than the assumed aperture size. This sample selection is relatively conservative compared to observational selections, however, we consider it a reasonable approximation.

We show the size--mass relationship for our biased sample of \textit{JWST}-observable galaxies in Fig.~\ref{fig:2d_size_mass} as a dashed black line. This line is within $\sim$0.1--0.2\,dex of the observational size--mass relationships, implying that observed size--mass relationships are strongly affected by the detectability of galaxies. 

\section{The size--mass relation across cosmic time}
\label{sec:The size--mass relation across cosmic time}

\subsection{Evolution of individual galaxies}
\label{sec:Evolution of individual galaxies}

The \thzoom simulations show a positive-size mass relation at high redshift and are consistent with \textit{JWST} observations (see Section~\ref{sec:Galaxy sizes}). This is in contrast to simulations such as \thesanone, IllustrisTNG, FLARES, and Bluetides, which have intrinsically negative size--mass relations at high redshift and must invoke heavily attenuated centers to recover observational results. 

\begin{figure} 
\centering
	\includegraphics[width=\columnwidth]{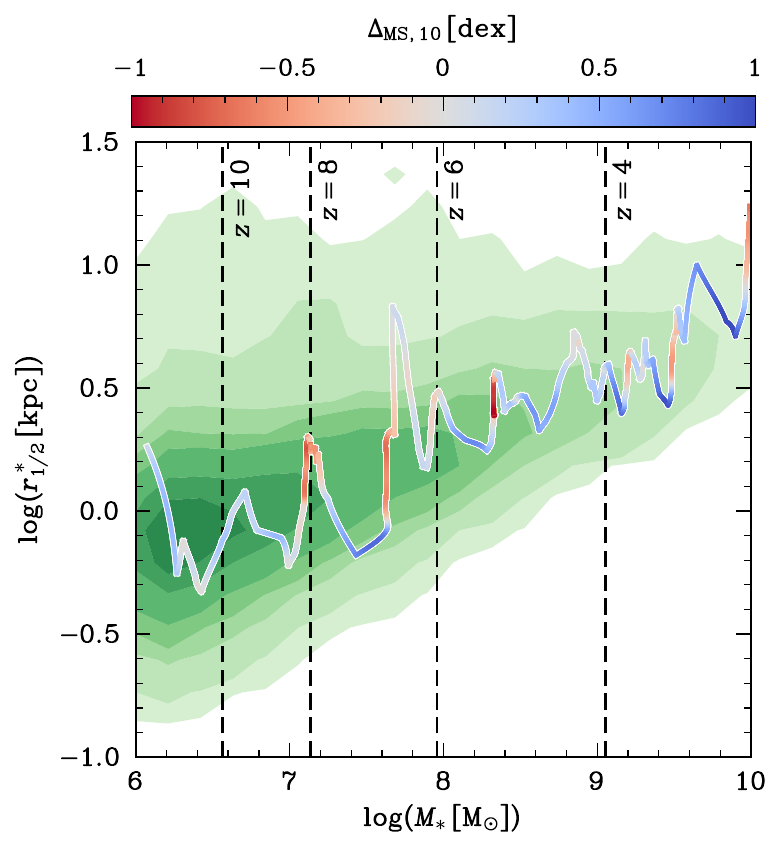}
    \caption{The stellar half-mass radii, $r^\ast_{1/2}$, of \thzoom galaxies as function of stellar mass. The contour plot shows the distribution of galaxies across $3<z<13$, with contours encompassing 20, 50, 80, 90, 95, 99, and 99.9$\%$ of the distribution. The line shows the size--mass evolution of an individual galaxy (m11.9, subhalo 0) and is coloured by its offset from the SFMS, $\Delta_{\mathrm{MS, 10}}$, where blue is above the SFMS and red is below. The dashed black lines show the mass of the galaxy at different redshifts. When the galaxy is heavily star-forming and lies above the SFMS, it goes through a phase of compaction where $r^\ast_{1/2}$ rapidly decreases due to strongly centrally concentrated star formation. Feedback and gas consumption cause the galaxy to quench inside out, leading to extended star formation and an increase in size in an expansion phase as the galaxy falls below the SFMS. 
    }
    \label{fig:stellar_mass_size_evo}
\end{figure}

We investigate the origin of the positive size--mass relation in Fig.~\ref{fig:stellar_mass_size_evo}, where we plot the size evolution of an individual galaxy as a function of mass. The size evolution of this galaxy is typical of galaxies in our simulation. It is immediately obvious that this galaxy does not evolve gently in size, instead showing dramatic fluctuations as it builds up mass (e.g. the half-mass size changes by factors of $2-3$ within $<50$\,Myr). To understand the cause of these fluctuations, we must understand how rapidly the galaxy is forming stars during its evolution. Simply coloring by SFR or sSFR is not informative as we cover a large mass and redshift range. A more useful characterization of how intensely a galaxy is forming stars relative to a typical galaxy at its particular stellar mass and redshift is the offset from the star-forming main sequence (SFMS), $\Delta_{\mathrm{MS}}$. We consider $\Delta_{\mathrm{MS, 10}}$, which is calculated as the logarithmic difference between SFR$_{10}$ and the \thzoom SFMS provided in \citet{McClymont:2025aa}. The line is coloured from blue to red according to whether the galaxy is above or below the SFMS, respectively. The size fluctuations are correlated with its position on the SFMS, such that the galaxy undergoes phases of compaction when it is above the SFMS, where $r^\ast_{1/2}$ decreases rapidly (on timescales comparable to the dynamical timescale), and the galaxy undergoes phases of expansion where $r^\ast_{1/2}$ increases rapidly. Similar, though less extreme, phases of size evolution have previously been identified in cosmological zoom-in simulations \citep{Zolotov:2015aa,Tacchella:2016aa,Emami:2021aa,Lapiner:2023aa}. 

\begin{figure} 
\centering
	\includegraphics[width=\columnwidth]{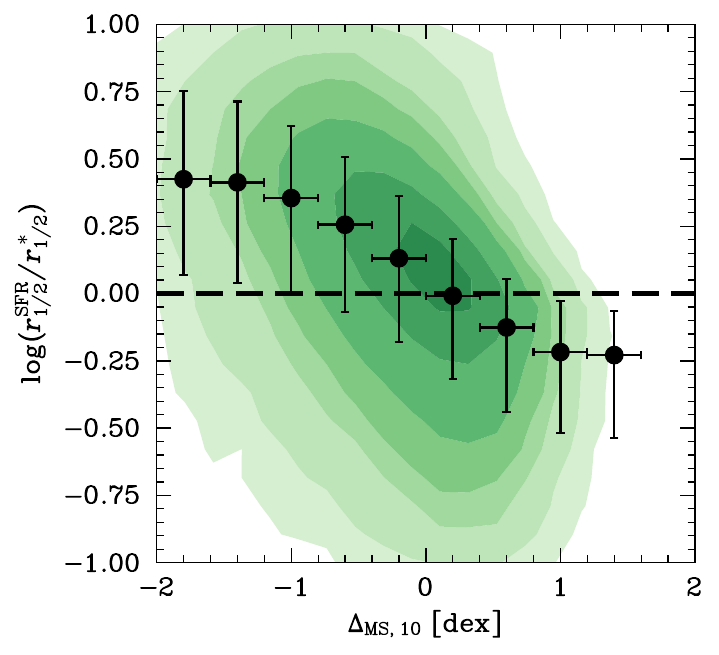}
    \caption{The half-mass radius of stars formed in the last 10\,Myr, $r^\mathrm{SFR}_{1/2}$, relative to the stellar half-mass radius, $r^\ast_{1/2}$, as a function of the offset from the SFMS, $\Delta_{\mathrm{MS, 10}}$. The contour plot shows the distribution of galaxies across $3<z<13$, with contours encompassing 20, 50, 80, 90, 95, 99, and 99.9$\%$ of the distribution. The black points and errorbars show the medians and $16^\text{th}$--$84^\text{th}$ percentile range, respectively. The black dashed line shows $r^{\mathrm{SFR}}_{\mathrm{1/2}}/r^{\ast}_{\mathrm{1/2}}=1$, on which galaxies fall if their star formation occurs on the same scale as the existing stellar population. Galaxies above the SFMS tend to form stars in a central starburst, causing compaction of the galaxy size. These starbursts tend to quench inside-out, meaning that most galaxies below the SFMS are expanding their size. As a galaxy oscillates about the SFMS, it also oscillates around the size--mass relation.}
    \label{fig:sfr_size_stellar_mass}
\end{figure}

\begin{figure*} 
\centering
	\includegraphics[width=\textwidth]{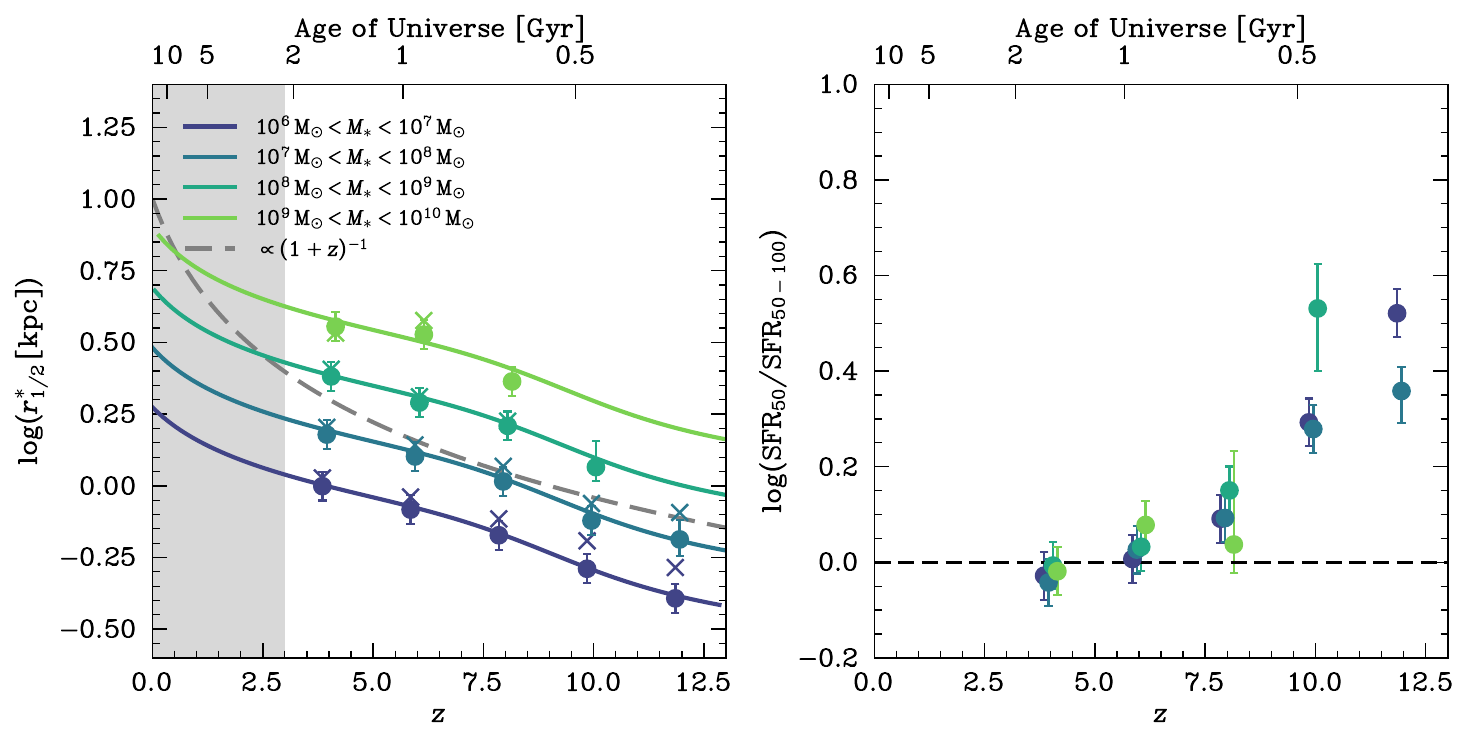}
    \caption{The redshift evolution of the size--mass relation. \textit{Left:} The stellar half-mass size, $r^\ast_{1/2}$, as a function of redshift. The dots and errorbars show the median and $16^\text{th}$--$84^\text{th}$ error on the median (from bootstrapping, 5\% error floor), respectively. Above $z\approx6$, the redshift evolution deviates from a simple $\propto(1+z)^\beta$ relation. The solid lines show our smoothly transitioning double power law fit. For context, we show a dashed gray line following a single power-law evolution with $\propto(1+z)^{-1}$. \textit{Right:} SFR$_{50}$/SFR$_{50-100}$ as a function of redshift. SFR$_{50}$/SFR$_{50-100}$ acts as an proxy for whether a galaxy is undergoing a burst or mini-quenching event. Above $z\approx6$ most galaxies are undergoing a starburst and therefore compacting in size, which causes the change in slope of the redshift evolution seen in the size--mass relation. The crosses in the left panel show the median $r^\ast_{1/2}$ for galaxies with $-0.1<\log(\mathrm{SFR}_{50}/\mathrm{SFR}_{50-100})<0.1$, $-0.5<\log(\mathrm{SFR}_{10}/\mathrm{SFR}_{10-50})<0.5$, and $-0.5<\log(\mathrm{SFR}_{10}/\mathrm{SFR}_{50-100})<0.5$, which helps to control for the fraction of galaxies with rising and falling SFHs. The crosses are closer to following a single power law across the entire redshift range.}
    \label{fig:redshift_evolution}
\end{figure*}

In order to investigate the causes of the compaction (size decrease) and expansion (size growth) phases, it is important to understand the relative sizes of recent star formation compared to the total distribution of stars. We investigate this on the population level in Fig.~\ref{fig:sfr_size_stellar_mass}, where we show the half-mass radius of stars formed in the last 10\,Myr, $r^\mathrm{SFR}_{1/2}$, relative to $r^\ast_{1/2}$ as a function of $\Delta_{\mathrm{MS, 10}}$. Galaxies above the SFMS tend to be forming stars on a smaller spatial scale than the existing stellar population, whereas galaxies below the SFMS tend to be forming stars on a larger spatial scale, which aligns with the evolution of the galaxy shown in Fig.~\ref{fig:stellar_mass_size_evo}. The episodes of compaction, where the galaxy size decreases, are driven by the inflow of gas to the center of galaxies which cause centrally concentrated starbursts. This does \textit{not} imply that the spatial distribution older stellar population has necessarily changed, simply that the new stars have shifted the radial mass profile to be more centrally concentrated. On the other hand, the episodes of expansion, where the galaxy size increases, are caused by the inside-out quenching of this star formation. By inside-out quenching we mean that the star formation in the center is suppressed, while the galaxy is still able to form stars on larger spatial scales. This inside-out quenching is likely caused by the strong feedback produced in the center of the galaxy during the initial starburst, as well as the depletion of gas as it is formed into stars. We note that the quenching we are referring to is more akin to the short-term SFR variability seen in high-redshift galaxies, rather than the long-term cessation of star formation in massive galaxies at lower redshift.

In this work, we refer to compaction and expansion of galaxy sizes on the relatively short timescales of starbursts \citep[$\sim30-100$\,Myr, see][]{McClymont:2025aa}. The mechanisms we describe as driving the evolution, central starbursts and inside-out quenching, also operate on similarly short timescales for our galaxies. However, size evolution has also been extensively studied at lower redshifts ($z<3$) and in more massive galaxies ($M_\ast\gtrsim10^{10}\,\mathrm{M}_\odot$), where these processes proceed over longer timescales. For example, \citet{Jain:2024aa} use \textit{HST} data to investigate galaxies $10^{9.8}\,\mathrm{M}_\odot<M_\ast<10^{11.5}\,\mathrm{M}_\odot$ at $0.5<z<2$ and find qualitatively similar correlations of size growth as a function of SFMS offset as we show in Fig.~\ref{fig:sfr_size_stellar_mass}.
Detailed investigations of statistical samples have not yet been performed at higher redshift, however, \citet{Baker:2024aa} found evidence for inside-out growth with \textit{JWST} observations of an individual galaxy.

The scatter in Fig.~\ref{fig:sfr_size_stellar_mass} also shows, however, that starburst-induced compaction and expansion are not the sole cause of changes in galaxy size, and there are certainly a myriad of other processes which can be important. For example, in addition to newly formed stars, sizes can also evolve due to the motion of the existing stellar population. Gas inflows and outflows can interact gravitationally with the stars, dragging them out to larger sizes \citep{El-Badry:2016aa}. We have not investigated this effect in detail; however, we believe this effect would be small compared to the spatial distribution of new stars. This is because galaxies at the masses and redshifts considered in this work form a significant fraction of their stellar mass, sometimes increasing by even an order of magnitude in a single burst (see Fig.~\ref{fig:stellar_mass_size_evo}), and as such the spatial extent of the newly formed stars is dominant over the older population. While the older stellar population is likely unimportant, one potentially important effect is if the stars are formed with significant velocity relative to the galaxy, such as if they form within an outflow. This may contribute to the expansion of galaxy sizes after star formation has ceased, although it is also unlikely to be dominant while star formation is ongoing.

The bursty nature of star formation in \thzoom gives rises to cyclical changes in the morphology of galaxies, allowing even the highest mass galaxies to have undergone expansion phases, rather than having been star-forming across their lifetime. It is therefore plausible that the positive high-redshift size--mass relation in the \thzoom simulations arises from compaction and expansion phases which are directly related to the bursty nature of star formation. This gives us a hint as to why other simulations instead find a negative size--mass relation; simulations such as FLARES, IllustrisTNG, and Bluetides use an unresolved ISM model and qualitatively different feedback prescriptions, which leads to much less bursty star formation \citep{Pillepich:2018aa,Lovell:2021aa,Feng:2016aa}. High-mass galaxies in the IllustrisTNG show incredibly rapid redshift evolution in their size at low-redshift in order to achieve a positive size--mass relation by $z=0$ \citep{Genel:2018aa,Tacchella:2019aa}. This growth is seen most strongly in galaxies that are quenched at $z=0$, which increase in size by over an order of magnitude as they fall below the SFMS from $z=3$ to $z=0$. This indicates very compact star formation until $z\approx3$, after which these massive galaxies quench and then grow in size due to galaxy-galaxy mergers. 

However, given other differences between the simulations, such as resolution, it is important to note that we cannot rule out other factors contributing to the positive size--mass relation. Additionally, due to the limited volume of the \thzoom simulations, we are not able to test how well our model reproduces the number densities and properties of the most massive, compact star-forming galaxies at high redshift. We also are not able to test our model across the full variety of environments. It is certain that our model will not correctly reproduce the morphologies for galaxies across all redshifts, masses, and environments, especially as we are missing key physical processes such as AGN feedback, but our results do at least indicate that burstiness is an important component in the regulation of galaxy sizes. This is interesting in the context of recent observational evidence for bursty star formation at high redshift \citep{Looser:2024aa,Looser:2025aa,Endsley:2024ab,Endsley:2025aa,Baker:2025aa,Trussler:2025aa,Witten:2025aa}.

The FIRE-2 simulations \citep{Ma:2018ab} also show bursty star formation, so we may expect that they also show a positive size--mass relation at high redshift. This tentatively appears to be the case, and FIRE-2 dwarf galaxies at $z=0$ show episodes of compaction and expansion during starbursts \citep{Emami:2021aa}, but unfortunately there is only one data point at $5<z<7$ around the turnover mass seen in other simulations, so it is difficult to draw concrete conclusions \citep{Ma:2018aa}.

\begin{table*}
    \centering
    \begin{tabular}{cccccccc}
        \hline
         \multicolumn{7}{|c|}{3D sizes} \\
        \hline
         & $s$ & $\beta$ & $\mu_1$ & $\mu_2$ & $z_b$ & $\Delta_b$\\
        $r^{\mathrm{\ast}}_{\mathrm{1/2}}$\,[kpc] & $3.9\pm0.2$ &  $0.191\pm0.002$ &  $-0.42\pm0.03$ &  $-0.64\pm0.09$ &  $9.1\pm0.2$ &  $2.8\pm0.5$ \\
        $r^{\mathrm{H\alpha}}_{\mathrm{1/2}}$\,[kpc] & $4.32\pm0.09$ &  $0.1412\pm0.0009$ &  $-0.30\pm0.01$ &  $-0.41\pm0.01$ &  $8.7\pm0.3$ &  $2.2\pm0.3$ \\
        $r^{\mathrm{UV}}_{\mathrm{1/2}}$\,[kpc] & $2.33\pm0.05$ &  $0.1965\pm0.0009$ &  $-0.16\pm0.01$ &  $-0.41\pm0.02$ &  $9.4\pm0.2$ &  $2.2\pm0.2$ \\
        $r^{\mathrm{opt}}_{\mathrm{1/2}}$\,[kpc] & $3.0\pm0.1$ &  $0.19\pm0.01$ &  $-0.33\pm0.03$ &  $-0.56\pm0.10$ &  $9.4\pm0.3$ &  $2.5\pm0.4$ \\
        \hline
        \multicolumn{7}{|c|}{2D sizes} \\
        \hline
         & $s$ & $\beta$ & $\mu_1$ & $\mu_2$ & $z_b$ & $\Delta_b$\\
        $R^{\mathrm{H\alpha}}_{\mathrm{eff}}$\,[kpc] & $2.59\pm0.07$ &  $0.147\pm0.001$ &  $-0.23\pm0.02$ &  $-0.35\pm0.02$ &  $9.2\pm0.4$ &  $2.2\pm0.5$ \\
        $R^{\mathrm{UV}}_{\mathrm{eff}}$\,[kpc] & $1.2\pm0.1$ &  $0.18\pm0.03$ &  $-0.07\pm0.02$ &  $-0.38\pm0.03$ &  $9.5\pm0.3$ &  $2.5\pm0.3$ \\
        $R^{\mathrm{opt}}_{\mathrm{eff}}$\,[kpc] & $1.46\pm0.08$ &  $0.18\pm0.02$ &  $-0.26\pm0.06$ &  $-0.51\pm0.02$ &  $9.7\pm0.3$ &  $2.7\pm0.4$ \\
        \hline
    \end{tabular}
    \caption{Our best fits to following a double power law redshift dependence and single power law mass dependence (Eq.~\ref{eq:size_mass_eq}). Our sample includes galaxies with $10^5\,\mathrm{M}_\odot<M_\ast<10^{11}\,\mathrm{M}_\odot$ at $3<z<13$. The relations were fit assuming a student-t error distribution and the quoted errors are a result of bootstrapping (resampling with replacement 1000 times).}
    \label{tab:size_mass_fits}
\end{table*}

\subsection{Redshift evolution}
\label{sec:Redshift evolution}

\subsubsection{Intrinsic redshift evolution}
\label{sec:Intrinsic redshift evolution}

The evolution of the size--mass relation with redshift has previously been parameterized as a function of redshift, $\propto(1+z)^\mu$ \citep{Shibuya:2015aa}, or as a function of the evolving Hubble constant $\propto H(z)^\mu$
\citep{van-der-Wel:2014aa}.

In the left panel of Fig.~\ref{fig:redshift_evolution} we show the size--mass relation as a function of redshift. The binned median values are shown as points, and it is immediately clear that the points do not follow prescriptions of $\propto(1+z)^\mu$ or $\propto H(z)^\mu$. A break is apparent at $z\approx6$, and so we instead fit a smoothly transitioning double power law for redshift and a single power law for stellar mass
\begin{equation}
\label{eq:size_mass_eq}
\mathrm{r_{1/2}}(\mathrm{M}_\ast,z)=\mathrm{s} \left( \frac{\mathrm{M}_\ast}{10^{8}~\mathrm{M}_\odot} \right)^\beta (1+z)^\mu~\mathrm{kpc}\,,
\end{equation}
where $\mu$ is defined by four parameters; the low-redshift power $\mu_1$, the high-redshift power $\mu_2$, the break redshift $z_b$, and the break smoothness $\Delta_b$. $\mu$ is calculated as
\begin{equation}
\label{eq:mu)eq}
\mu = \mu_1 + \frac{1}{2}\left(1+ \tanh\left( \frac{{z-z_b}}{\Delta_b}\right)\right) \left(\mu_2-\mu_1 \right)\,.
\end{equation}
To carry out this fit we minimize a negative log-likelihood defined by a student's t distribution, with a shape parameter $\nu$ and scale parameter $\sigma$. We found that this fit the peak of the size--mass distribution more effectively than a Gaussian likelihood, as it could better handle the tail of extremely extended galaxies. For each fit there are 8 parameters $s$, $\beta$, $\mu_1$, $\mu_2$, $z_b$, $\Delta_b$, $\nu$, and $\sigma$. We note that we are fitting to the entire distribution of galaxy sizes, and that the median points shown in Fig.~\ref{fig:redshift_evolution} are simply illustrative.

The four solid lines in the left panel of Fig.~\ref{fig:redshift_evolution} show our best fit for $r^\ast_{1/2}$ with Eq.~\ref{eq:size_mass_eq} for different stellar masses. The double power law form well fits the evolving peak of the size--mass relation as shown by the medians. The double power law form also well fits the half-light sizes, including their 2D projections. We show the parameters for all fits in Tab.~\ref{tab:size_mass_fits}. We note that if we instead use a single power-law fit, the total redshift evolution tends to be dominated by $\mu_2$. For example, for $r^{\mathrm{\ast}}_{\mathrm{1/2}}$, the single-power law redshift evolution is given by $\mu=-0.626\pm0.004$.
This redshift evolution is shallower than the evolution found in most observational and theoretical studies, which tends to be $-2\lesssim\mu\lesssim-1$ \citep[e.g.,][]{Oesch:2010aa,Mosleh:2012aa,Roper:2022aa,Punyasheel:2025aa}, although similar values have been found in the Bluetides simulations \citep[$\mu=-0.662\pm0.009$][]{Marshall:2022aa} and some high-redshift observations with \textit{HST} \citep[$\mu=-0.72\pm0.12$][]{Holwerda:2015aa} and \textit{JWST} \citep[$\mu=-0.81$][]{Allen:2025aa}. One possible reason for our shallower overall redshift evolution, at least for observed sizes, could be that dust does not have a strong impact on observed sizes in our simulations. An increasing impact of dust, which is generally expected to cause observed sizes to be larger than intrinsic sizes due to concentrated dust \citep[e.g.,][]{Roper:2022aa}, with redshift would cause a steeper observed redshift evolution.

\subsubsection{Origin of the double power law redshift evolution}
\label{sec:Cause of the double power law redshift evolution}

That the galaxies in our simulation show double power law redshift dependence indicates that galaxy sizes in the early Universe may be governed by different physical processes than at low redshift. There are various redshift-dependent effects that could impact galaxy sizes, such as reionization \citep{Shen:2024aa} and an increasing merger rate \citep{Duan:2025aa,Puskas:2025aa}. However, given the dramatic impact that bursty star formation has on galaxy sizes (as demonstrated in Section~\ref{sec:Evolution of individual galaxies}), we instead favor a simpler explanation. In the right panel of Fig.~\ref{fig:redshift_evolution} we show the evolution of $\mathrm{SFR}_{50}/\mathrm{SFR}_{50-100}$ as a function of redshift. $\mathrm{SFR}_{50}/\mathrm{SFR}_{50-100}$ acts as a rough proxy for galaxies beginning a starburst or quenching phase. The relatively long averaging timescales of 50\,Myr and 100\,Myr are appropriate for our purposes because they are less sensitive to short-term noise and therefore better trace the timescales more relevant to changes in the stellar morphology. Galaxies of all mass ranges have increasing median $\mathrm{SFR}_{50}/\mathrm{SFR}_{50-100}$ with redshift. This shows that most galaxies at high redshift have a rising SFHs, and therefore are undergoing a compaction phase, and so are smaller than expected. Below $z\approx6$, an equilibrium is established, with a population of galaxies with both rising and falling SFRs.

The crosses in the left panel of Fig.~\ref{fig:redshift_evolution} show the median values for galaxies with $-0.1<\log(\mathrm{SFR}_{50}/\mathrm{SFR}_{50-100})<0.1$. To further ensure the selected galaxies are in relatively stable periods of star formation, we also require that $-0.5<\log(\mathrm{SFR}_{10}/\mathrm{SFR}_{10-50})<0.5$ and $-0.5<\log(\mathrm{SFR}_{10}/\mathrm{SFR}_{50-100})<0.5$. This acts to somewhat normalize the fraction of galaxies bursting and quenching across the redshift range, albeit it is not a perfect correction. The crosses lie above the total median points at high redshifts, demonstrating that the double power law form is due, at least in part, to the large fraction of galaxies compacting during a starburst above $z\approx6$. However, we must caution that the selection for stable SFHs which we have made is not perfect and is limited in its aggressiveness due to our sample size. Additionally, the \thzoom simulations do not represent unbiased selections of galaxies, and so we may be biased by environmental effects. A larger sample of simulated galaxies across a representative selection of environments would help confirm this evolution, and further establish the importance of other effects, such as mergers, in governing galaxy sizes.

\subsubsection{Comparison with observed evolution}
\label{sec:Comparison with observed evolution}

\begin{figure} 
\centering
	\includegraphics[width=\columnwidth]{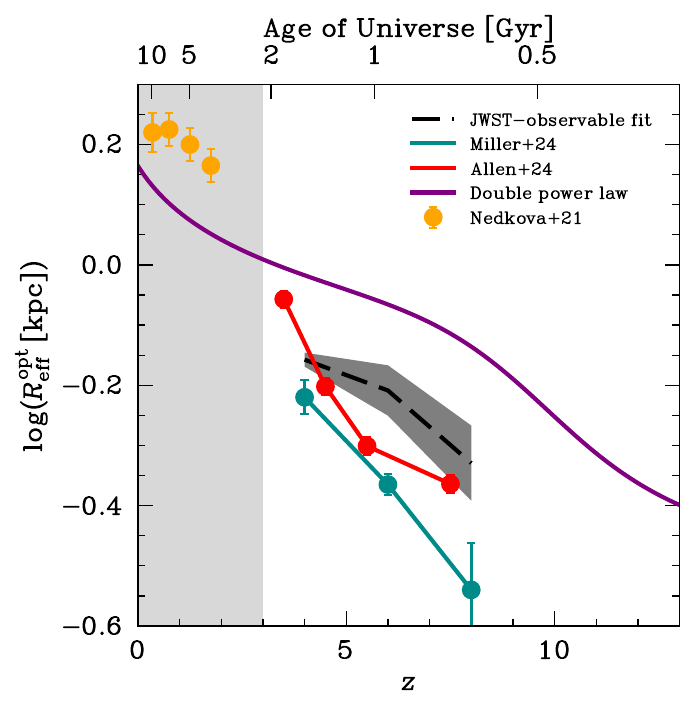}
    \caption{The redshift evolution of optical galaxy sizes. The lines show the predicted size for a $M_\ast=10^{8}\,\mathrm{M}_\odot$ galaxy from our best fit double power law to $R^{\mathrm{opt}}_{\mathrm{eff}}$ (purple) and for our redshift-binned fit to $R^{\mathrm{opt}}_{\mathrm{eff}}$ of our \textit{JWST}-observable sample (dashed black). The dark-shaded region shows a 1$\sigma$ error from our fit. We also show low-redshift data from \citet{Nedkova:2021aa} and our extrapolated relation. The light-shaded region shows the region where our fits are extrapolated ($z<3$). Our \textit{JWST}-biased sample is considerably closer to \textit{JWST} observations from \citet{Miller:2025aa} and \citet{Allen:2025aa} than our intrinsic data.}
    \label{fig:observed_redshift_evolution}
\end{figure}

The double power law redshift dependence which we find in this work is, in principle, observable. However, as demonstrated in Section~\ref{sec:Comparison to observations}, observational biases can have a strong impact on derived size--mass relations. Our findings imply that this biasing is particularly acute because quenching galaxies are less observable due to two effects; they are intrinsically dim due to their declining SFHs, and they are extended due to inside-out quenching causing suppressed central star formation. These effects in combination generate a population of galaxies with relatively low surface brightness which are therefore difficult to observe.

In Fig.~\ref{fig:observed_redshift_evolution} we show the redshift evolution of optical size for a $M_\ast=10^{8}\,\mathrm{M}_\odot$ galaxy lying on our best fit relations. The solid purple line shows the double power law best fit for $R^{\mathrm{opt}}_{\mathrm{eff}}$. We also show our observationally biased fit in redshift bins, as described in Section~\ref{sec:Comparison to observations}. The observationally biased fit is consistent within $\sim$0.1\,dex with optical size--mass relations derived from \textit{JWST} observations \citep{Miller:2025aa,Allen:2025aa}. This suggests that the strong redshift evolution implied by the \textit{JWST}-based fits may be partially due to selection effects. We also show low redshift data from the size--mass relations of star-forming galaxies in \citet{Nedkova:2021aa} to place our extrapolated relation in context.

\section{Extended nebular emission}
\label{sec:Extended nebular emission}

\begin{figure*} 
\centering
	\includegraphics[width=\textwidth]{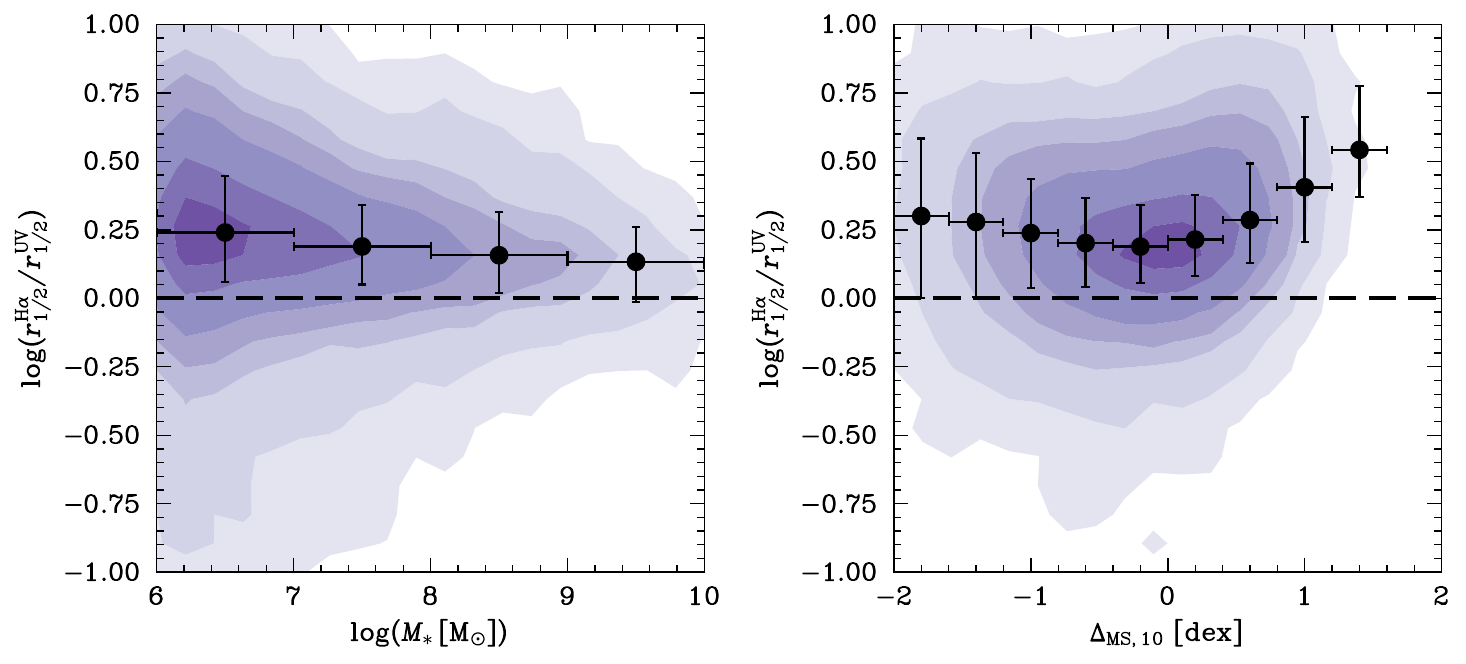}
    \caption{The ratio of intrinsic H$\alpha$ to UV sizes, $\log(r^{\mathrm{H\alpha}}_{\mathrm{1/2}}/r^{\mathrm{UV}}_{\mathrm{1/2}})$, against stellar mass (left) and SFMS offset, $\Delta_{\mathrm{MS, 10}}$ (right). The contour plots show the distribution of galaxies across $3<z<13$, with contours encompassing 20, 50, 80, 90, 95, 99, and 99.9$\%$ of the distribution. The black points and errorbars show the medians and $16^\text{th}$--$84^\text{th}$ percentile range respectively. The black dashed line shows $r^{\mathrm{H\alpha}}_{\mathrm{1/2}}/r^{\mathrm{UV}}_{\mathrm{1/2}}=1$. Across the entire sample, the median $r^{\mathrm{H\alpha}}_{\mathrm{1/2}}/r^{\mathrm{UV}}_{\mathrm{1/2}}=1.7^{+1.0}_{-0.5}$, where the errors are the $16^\text{th}$--$84^\text{th}$ percentile range. Lower mass galaxies tend to have more extended H$\alpha$ emission. This can be explained as the byproduct of a more fundamental dependence on the $\Delta_{\mathrm{MS, 10}}$, where $\log(r^{\mathrm{H\alpha}}_{\mathrm{1/2}}/r^{\mathrm{UV}}_{\mathrm{1/2}})$ increases as galaxies are further offset from the SFMS. Lower mass galaxies show higher scatter around the SFMS, thereby causing the observed trend with mass \citep{McClymont:2025aa}.}
    \label{fig:ha_uv_size_ratio}
\end{figure*}

Recent studies with \textit{JWST} find that high redshift galaxies show generally extended nebular sizes (Villanueva et al. in prep., Danhaive et al. in prep.) and radially increasing emission line EWs \citep{Nelson:2024aa,Matharu:2024aa}. It has been claimed that this is a result of inside-out growth, where star formation occurs at successively larger radii as a galaxy builds up its mass \citep{Nelson:2024aa,Matharu:2024aa}. Our simulated galaxies also tend to show extended H$\alpha$ emission relative to the continuum, in agreement with the observations. However, as discussed in Section~\ref{sec:Evolution of individual galaxies}, galaxy sizes do not evolve in a simple and steady manner, instead showing phases of compaction and expansion. In fact, galaxies above the SFMS, which show the strongest nebular emission, tend to form their stars in a central starburst. Already we can see strong tension with the interpretation of extended nebular sizes as evidence of inside-out growth. Throughout this section, we will investigate the causes of extended nebular emission.

\subsection{Which galaxies show extended nebular emission?}
\label{sec:Which galaxies show extended nebular emission?}

It is first informative to understand which population of galaxies show extended nebular sizes, and so we begin our investigation in Fig.~\ref{fig:ha_uv_size_ratio}, where we show $\log(r^{\mathrm{H\alpha}}_{\mathrm{1/2}}/r^{\mathrm{UV}}_{\mathrm{1/2}})$ against stellar mass in the left panel. There is a clear, albeit weak, trend of increasing $r^{\mathrm{H\alpha}}_{\mathrm{1/2}}/r^{\mathrm{UV}}_{\mathrm{1/2}}$ with decreasing stellar mass. Although this trend is weak, a linear fit of $\log(r^{\mathrm{H\alpha}}_{\mathrm{1/2}}/r^{\mathrm{UV}}_{\mathrm{1/2}})$ as a function of $\log M_\ast$ shows that it is statistically significant, with a slope of $-0.0472\pm0.0009$ (error from bootstrapping). However, there is no obvious fundamental link between $r^{\mathrm{H\alpha}}_{\mathrm{1/2}}/r^{\mathrm{UV}}_{\mathrm{1/2}}$ and stellar mass. The dominant source of both UV and H$\alpha$ emission in star-forming galaxies is recent star-formation, and therefore it is reasonable to expect that there is a more fundamental relationship between $r^{\mathrm{H\alpha}}_{\mathrm{1/2}}/r^{\mathrm{UV}}_{\mathrm{1/2}}$ and recent star formation.

In the right panel of Fig.~\ref{fig:ha_uv_size_ratio} we show $\log(r^{\mathrm{H\alpha}}_{\mathrm{1/2}}/r^{\mathrm{UV}}_{\mathrm{1/2}})$ against $\Delta_{\mathrm{MS, 10}}$, where we can see a clear relationship. $r^{\mathrm{H\alpha}}_{\mathrm{1/2}}/r^{\mathrm{UV}}_{\mathrm{1/2}}$ is at a minimum when a galaxy lies on the SFMS, where $\Delta_{\mathrm{MS, 10}}\approx0$. There is a rapid increase in $r^{\mathrm{H\alpha}}_{\mathrm{1/2}}/r^{\mathrm{UV}}_{\mathrm{1/2}}$ with increasing $\Delta_{\mathrm{MS, 10}}$ above the SFMS, and a more mild increase in $r^{\mathrm{H\alpha}}_{\mathrm{1/2}}/r^{\mathrm{UV}}_{\mathrm{1/2}}$ with decreasing $\Delta_{\mathrm{MS, 10}}$ below the SFMS. By interpreting this relationship as the more fundamental dependence, we can understand the origin of the weaker trend with stellar mass. Lower mass galaxies show a larger scatter around the SFMS \citep{McClymont:2025aa}, and therefore low mass galaxies have a larger average absolute offset from the SFMS. The trend of increasing $r^{\mathrm{H\alpha}}_{\mathrm{1/2}}/r^{\mathrm{UV}}_{\mathrm{1/2}}$ with decreasing stellar mass can simply be explained as a byproduct of the more fundamental relationship between $r^{\mathrm{H\alpha}}_{\mathrm{1/2}}/r^{\mathrm{UV}}_{\mathrm{1/2}}$ and $\Delta_{\mathrm{MS, 10}}$.

\subsection{Extended LyC emission}
\label{sec:Extended LyC emission?}

The dominant source of H$\alpha$ emission in star-forming galaxies is recombinations following the ionization of \HI by the copious amount of LyC photons produced by recently formed stars. UV emission is similarly produced by young stars, although it traces a longer averaging timescale of $\sim$24\,Myr compared to $\sim$8\,Myr for H$\alpha$ \citep{McClymont:2025aa}. If one assumes that LyC photons are absorbed locally to their source star, the H$\alpha$ and UV morphology can be used to trace the spatial distribution of these two stellar populations. This is the basis for the interpretation of observed nebular sizes as evidence for inside-out growth \citep{Nelson:2024aa,Matharu:2024aa}.

We are able to trace the emission of LyC photons directly in our simulation and therefore we can directly test extended LyC sizes as the cause of the extended H$\alpha$ sizes. In the left panel of Fig.~\ref{fig:triple_size_ratio_plots} we show the LyC to H$\alpha$ size ratio as a function of $\Delta_{\mathrm{MS, 10}}$. LyC sizes are systemically $\sim$0.3\,dex smaller than H$\alpha$ sizes, and the discrepancy increases for galaxies above the SFMS, reaching $\sim$0.5\,dex at $\Delta_{\mathrm{MS, 10}}>1$.

The discrepancy between the LyC and H$\alpha$ sizes remains consistently $\sim$0.3\,dex for galaxies below the SFMS rather than increasing, as one might expect due to the mildly increasing $r^{\mathrm{H\alpha}}_{\mathrm{1/2}}/r^{\mathrm{UV}}_{\mathrm{1/2}}$ below the SFMS. This is because the LyC sizes also show a mild increase relative to UV sizes as $\Delta_{\mathrm{MS, 10}}$ decreases for galaxies below the SFMS. This is due to the expansion phase as a galaxy quenches inside-out. LyC emission may therefore at least partly explain why galaxies below the SFMS show extended nebular sizes, but even in this case it does not appear to be the dominant factor. Additionally, this explanation completely fails to explain the extended nebular sizes for heavily star-forming galaxies.

\subsection{Extreme ionizing conditions}
\label{sec:Extreme ionizing conditions}

\begin{figure*} 
\centering
	\includegraphics[width=\textwidth]{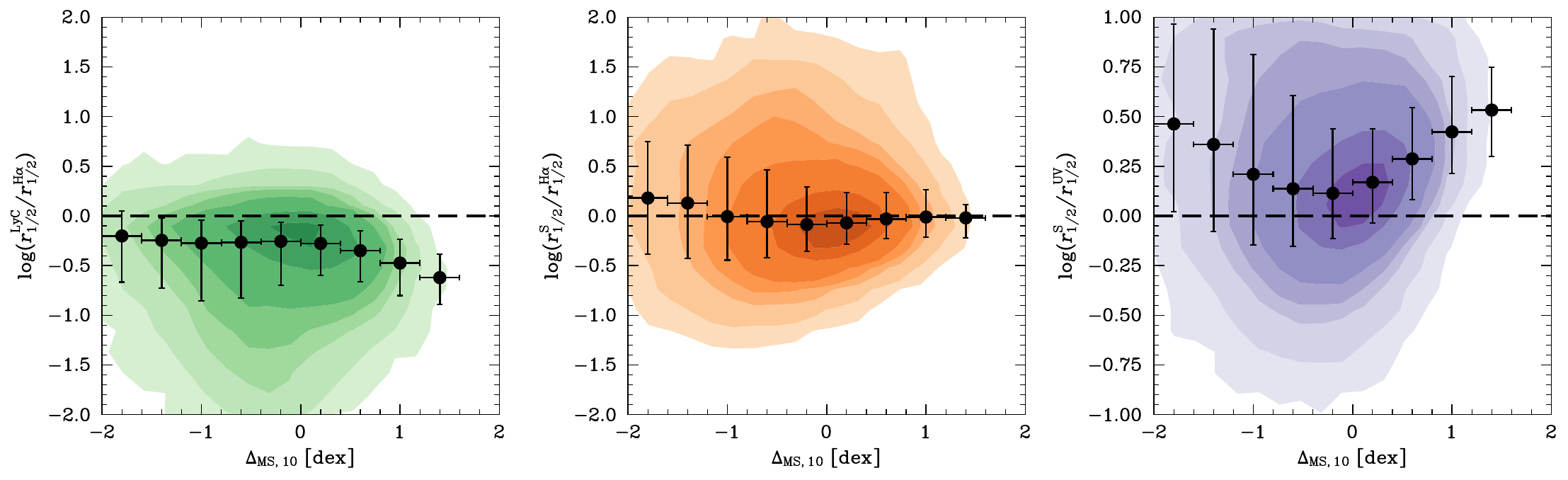}
    \caption{The origin of extended nebular emission.
    The contour plots show the distribution of galaxies across $3<z<13$, with contours encompassing 20, 50, 80, 90, 95, 99, and 99.9$\%$ of the distribution. The black points and errorbars show the medians and $16^\text{th}$--$84^\text{th}$ percentile range respectively. The black dashed line shows where the ratios equal unity.
    \textit{Left:} The ratio of intrinsic LyC to H$\alpha$ sizes, $\log(r^{\mathrm{LyC}}_{\mathrm{1/2}}/r^{\mathrm{H\alpha}}_{\mathrm{1/2}})$, against SFMS offset, $\Delta_{\mathrm{MS, 10}}$. LyC sizes generally under-predict H$\alpha$ sizes by 0.2--0.3\,dex, indicating that LyC and H$\alpha$ emission is spatially disconnected in high-redshift galaxies. \textit{Center:} The ratio of Strömgren half-light radius to H$\alpha$ size, $\log(r^{\mathrm{S}}_{1/2}/r^{\mathrm{H\alpha}}_{\mathrm{1/2}})$, against $\Delta_{\mathrm{MS, 10}}$. H$\alpha$ sizes are fairly well approximated by the estimated Strömgren half-light radius across the SFMS. Starburst galaxies produce such significant LyC emission that their Strömgren radius extends well beyond the galaxy itself, causing extended H$\alpha$ emission. For galaxies below the SFMS, the average density of the ISM and CGM tends to be lower, which correspondingly increases the Strömgren radius, causing an uptick at $\Delta_{\mathrm{MS, 10}}<-1$. \textit{Right:} The ratio of Strömgren radius to UV sizes, $\log(r^{\mathrm{S}}_{1/2}/r^{\mathrm{UV}}_{\mathrm{1/2}})$, against $\Delta_{\mathrm{MS, 10}}$. This simple model for H$\alpha$ sizes is successful at qualitatively reproducing the $r^{\mathrm{H\alpha}}_{\mathrm{1/2}}/r^{\mathrm{UV}}_{\mathrm{1/2}}$ trend as a function of $\Delta_{\mathrm{MS, 10}}$, which leads us to conclude that it captures the dominant causes of extended H$\alpha$ sizes. There are numerous effects not accounted for in our simple model (e.g. realistic geometry, non-equilibrium ionization states, outflows) which affect the H$\alpha$ sizes at a subdominant level.
    }
    \label{fig:triple_size_ratio_plots}
\end{figure*}

We have shown that H$\alpha$ sizes are consistently $\sim$0.3\,dex larger than LyC sizes and the discrepancy is most extreme for heavily star forming galaxies. This implies that H$\alpha$ emission is spatially disconnected from the LyC sources which ionized the gas, which we can also see in the RBG images shown in Fig.~\ref{fig:thumbnail_images_6x6}. The spatial disconnection of H$\alpha$ emission from LyC sources is not unprecedented, LyC photons leaking from young stars are thought to be the cause of diffuse ionized gas in local star-forming galaxies \citep{Levy:2019aa,Belfiore:2022aa,McClymont:2024aa}. The star-formation in these local galaxies is tame relative to the high-redshift galaxies considered in this work, which means that the density of ionizing photons is also lower than the galaxies we consider. In addition, the galaxies in this work are low-mass and small in size relative to local star-forming galaxies, which means that radiative transfer effects can become relatively more important; a LyC mean free path of 1\,kpc would have little effect on the nebular size of a 15\,kpc galaxy, but a dramatic effect on a 0.5\,kpc galaxy.

We can investigate this by considering a simple model for a nebula around an ionizing source: the Strömgren sphere \citep{Stromgren:1939aa}. This model assumes a constant density and temperature gas is ionized by a point-source at the center of the sphere and that the gas is completely ionized until the Strömgren radius is reached, where the ionizing photons are depleted. We approximate each galaxy as a point source at the center of a Strömgren sphere and calculate the Strömgren radius. The point-source emits the same rate of ionizng photons as the galaxy and the density is the average hydrogen density within twice the stellar half-mass radius. We set a density floor of $n_\mathrm{H}=10^{-3}\,\mathrm{cm^{-3}}$, which acts to prevent spuriously low densities. This floor impacts $4.9\%$ of galaxies at $-2<\Delta_{\mathrm{MS, 10}}<2$, which are mostly lower mass galaxies $\Delta_{\mathrm{MS, 10}}<-1$ with few gas particles within twice the stellar half-mass radius. We only consider the ionization of hydrogen. We also assume a constant temperature of 15\,000\,K when calculating the recombination rate, for which we use the Case B rate from \citet{Hui:1997aa}. As we are interested in the H$\alpha$ half-light size, we calculate the Strömgren half-light radius as $r^{\mathrm{S}}_{1/2}=0.8r^{\mathrm{S}}$, which is equivalent to the half-mass radius for a uniform sphere. We stress that this is a toy model to investigate the plausibility of this scenario in causing extended H$\alpha$ sizes, and that both the simulation and reality are much more complex.

\begin{figure*} 
\centering
	\includegraphics[width=\textwidth]{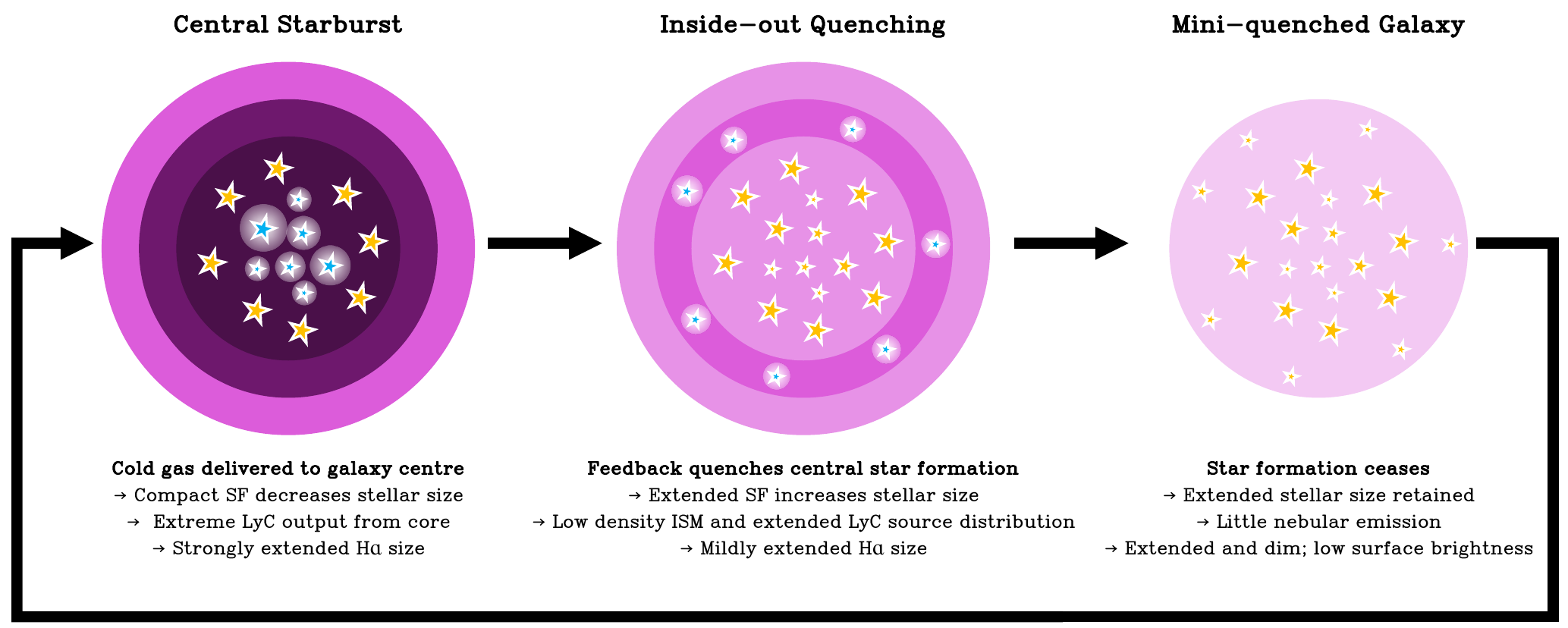}
    \caption{The evolution of galaxy sizes throughout a starburst. Blue represents young star, orange represents older stars, and the intensity of H$\alpha$ emission is shown by the darkness of the purple circles. Starbursts tend to arise from the rapid supply of cold gas to the center of a galaxy, which causes strongly concentrated star formation and the stellar size to compact drastically. The nebular emission is generally strongly extended during this period due to the extreme LyC emission from the compact star-forming core creating a large Strömgren sphere. Stellar feedback causes star formation to quench from the inside out, leading to extended star formation and an increase in the stellar size. H$\alpha$ tends to be mildly extended during this phase due to a low ISM density extending the Strömgren sphere. Eventually, all star formation ceases and the galaxy retains its extended stellar size. These ``mini-quenched'' galaxies show little nebular emission. They also have low continuum surface brightness due to their extended nature and dim, older stellar population.
    }
    \label{fig:size_diagram}
\end{figure*}

In the central panel of Fig.~\ref{fig:triple_size_ratio_plots} we show the Strömgren half-light radius, $r^{\mathrm{S}}_{1/2}$, relative to $r^{\mathrm{H\alpha}}_{\mathrm{1/2}}$ across the SFMS. The degree of agreement is striking, with all median points lying within $\pm$$\sim$0.2\,dex. This agreement across the whole of the SFMS requires that the Strömgren half-light radius increases relative to the UV size for increasing $\Delta_{\mathrm{MS, 10}}$ above the SFMS \textit{and} for decreasing $\Delta_{\mathrm{MS, 10}}$ below the SFMS, which is shown in the right panel of Fig.~\ref{fig:triple_size_ratio_plots}. The trend above the SFMS can be simply explained by the copious LyC emission in such heavily star-forming galaxies increasing their Strömgren radii. However, galaxies below the SFMS produce fewer LyC photons as they have fewer young stars, and so we may expect that $r^{\mathrm{S}}_{1/2}$ would continue to decrease below the SFMS. However, decreasing ISM densities below the SFMS are sufficient to invert this trend. In Fig.~\ref{fig:size_diagram} we show a schematic representation of how the stellar and nebular sizes of a galaxy evolve throughout a starburst.

We note that there is some freedom in the modeling to change the normalization of $r^{\mathrm{S}}_{1/2}$ across the whole population by $\sim$0.2\,dex for reasonable choices of temperature (7\,000--30\,000\,K). However, this does not impact the shape of the relation with $\Delta_{\mathrm{MS, 10}}$ and so the qualitative agreement of the trends is independent of these parameter choices.

The significant scatter shows that while our model generally does well at reproducing population-scale trends, there is uncertainty on the level of individual galaxies. These shortcomings are likely due to a large number of complex factors which are accounted for in our simulation and post-processing methods, but are not reflected in our simple model. We have assumed that the gas is uniformly distributed, when we know that gas is actually denser around sites of recent star formation and is generally heavily clumpy. This would likely dampen the trends as it would reduce the mean-free path of the LyC photons. Additionally, the most extreme disagreements of $\log(r^{\mathrm{S}}_{1/2}/r^{\mathrm{H\alpha}}_{\mathrm{1/2}})\gg2$ tend to result from extremely low densities, approaching the density floor at $n_\mathrm{H}=10^{-3}\,\mathrm{cm^{-3}}$. This indicates that the density within twice the stellar half-mass radius is not necessarily appropriate for these galaxies.

Another factor is that the fraction of ionized hydrogen in our simulation is calculated on-the-fly and is not necessarily in ionization equilibrium. This is at odds with the implicit assumption of ionization equilibrium when calculating a Strömgren radius. In the relatively low densities of the CGM, the recombination time of ionized gas can become significant, which may cause H$\alpha$ emission to remain extended for tens of Myr after the intensity of ionizing radiation has reduced \citep{McCallum:2024aa}. While this is unlikely to dominate the total flux, it may impact the sizes.

Additionally, H$\alpha$ emission from outflowing gas is likely to cause extended sizes for some galaxies, however given the success at reproducing H$\alpha$ sizes purely considering radiative transfer of LyC photons, we do not think that outflows act to extend H$\alpha$ sizes significantly on a population level. Outflows do also indirectly contribute to extended H$\alpha$ sizes because they tend to shortly follow bursts of star formation and clear the ISM of a mini-quenched or lulling galaxy \citep{McClymont:2025aa}, and therefore are key in extending the Strömgren radii of galaxies below the SFMS.

Overall, these results suggest that the extended nebular sizes in high–redshift galaxies can be largely understood in terms of extreme LyC output in starburst galaxies ionizing gas well beyond the ISM scale. This implies that extended nebular sizes are strongly related to LyC escape, at least on a galactic scale. Developing robust LyC escape diagnostics is vital to understand which population of galaxies was responsible for reionizing the Universe \citep{Naidu:2020aa} and are widely used in the study of high-redshift galaxies \citep{Witten:2023aa,Gazagnes:2025aa,Mascia:2025aa}. Unfortunately LyC diagnostics remain heavily uncertain \citep{Chisholm:2018aa, Choustikov:2024aa,Choustikov:2025aa}. Extended H$\alpha$ sizes may therefore prove a useful diagnostic, particularly as these galaxies are undergoing starbursts and therefore producing copious LyC emission \citep{Endsley:2023aa,Simmonds:2023aa,Simmonds:2024aa,Simmonds:2024ab,Laseter:2025aa}. Understanding the relationships between extended nebular sizes and other galaxy properties may also provide insight as to whether this mode of leakage can also help explain the prevalence of Lyman-$\alpha$ emitters \citep[LAEs; ][]{Saxena:2024aa,Witten:2024aa,Tang:2024ab,Goovaerts:2024aa,Kumari:2024aa,Jones:2025aa} and anomalous Balmer emitters \citep[ABEs; ][]{McClymont:2025ad} at high redshift. We plan to explore this in future work focused on understanding LyC escape diagnostics.

\section{Summary and Conclusions}
\label{sec:Summary and Conclusions}

In this work, we have explored the evolution of galaxy sizes in the early Universe with the \thzoom simulations. These simulations are high-resolution zoom-ins of the original \thesan simulations, and provide the ideal laboratory for studying galaxy sizes during the epoch of reionization due to the self-consistent implementation of the ionizing radiation background. Combined with a resolved model of the interstellar medium (ISM), this allows us to accurately capture the behavior of high-redshift galaxies which is relevant to their size evolution, such as bursty star formation. We have also used the post-processing code \textsc{colt} to generate synthetic H$\alpha$ and UV observables which has allowed us to directly compare our results to observational studies. In our analysis, we focused on galaxies across a wide mass and redshift range of $10^6\,\mathrm{M}_\odot<M_\ast<10^{10}\,\mathrm{M}_\odot$ and $3<z<13$, respectively. Our key findings are as follows:

\begin{itemize} 
\item \textbf{Consistency with observations:} We reproduce the observed size--mass relation from \textit{JWST} data when observational biases, such as selection effects due to survey depth, are accounted for. By considering galaxies observable in typical surveys, our biased size--mass relations lie within $\sim$0.1\,dex of those measured with \textit{JWST}. 

\item \textbf{Relationship between sizes and the burst--quench cycle:} We demonstrate that galaxy size evolution is strongly tied to the bursty nature of high-redshift star formation. Star formation in galaxies above the star-forming main sequence (SFMS) is centrally concentrated, causing them to undergo morphological compaction. These starbursts quench inside out, causing their sizes to increase as star formation declines. These alternating phases of compaction and expansion cause galaxies to oscillate about the size--mass relation.

\item \textbf{Positive high-redshift size--mass relation:} We find a positive size--mass relation at high redshift. This is in contrast to simulations such as IllustrisTNG and FLARES, which have intrinsically negative size--mass relations at high redshift and rely on concentrated dust attenuation in massive galaxies to reproduce the observed size--mass relation. We attribute this to the lack of bursty star formation in these simulations, which means they do not reproduce the same compaction and expansion phases.

\item \textbf{Double power law redshift evolution:} The intrinsic evolution of galaxy sizes with redshift in the \thzoom simulations follows a double power law, with a break above $z\approx6$. This more rapid size evolution with redshift may be because most galaxies at high redshift have rising SFHs, and are therefore in a compaction phase. Below $z\approx6$, galaxies have established an equilibrium of rising and falling SFRs and so sizes follow a simple power law evolution.

\item \textbf{Extended nebular sizes:} Galaxies in our simulation show extended H$\alpha$ sizes relative to the UV continuum, with a median of $r^{\mathrm{H\alpha}}_{\mathrm{1/2}}/r^{\mathrm{UV}}_{\mathrm{1/2}}=1.7^{+1.0}_{-0.5}$, where the errors are the $16^\text{th}$--$84^\text{th}$ percentile range. This effect becomes more pronounced as the offset from the SFMS increases. Galaxies above the SFMS produce such extreme LyC emission that their Strömgren radii lie well outside their UV size. Galaxies below the SFMS have low ISM densities, increasing their Strömgren radius despite low LyC luminosities.

\end{itemize}

Overall, our results demonstrate that resolving detailed ISM physics and bursty star formation is crucial for accurately reproducing the observed evolution of galaxy sizes in the early Universe. In particular, our work suggests that a positive high redshift size--mass relation, double power law redshift evolution, and extended nebular emission are natural consequences of bursty star formation.

Finally, we note that the \thzoom simulations do not include a live black hole (BH) model. That we have found high-redshift galaxies in the \thzoom model to be extremely bursty \citep{McClymont:2025aa}, and that these bursts tend to be centrally concentrated, naturally begs the question as to how this extreme mode of star formation relates to the growth of BHs. This is especially interesting given the compact nature of observed quasar host galaxies \citep{Ding:2022aa} and apparent abundance of overmassive (relative to their host galaxies) black holes at high redshift \citep[e.g.,][]{Harikane:2023aa,Maiolino:2024aa,Maiolino:2024ac,Juodzbalis:2024ab,Natarajan:2024aa,Geris:2025aa,McClymont:2025ac}. We anticipate future work integrating a BH model into the \thzoom simulations to study the co-evolution of galaxies and their BHs in the context of bursty star formation and central starbusts in the early Universe.

\section*{Acknowledgements}

The authors are grateful to the anonymous referee for providing constructive feedback which improved the manuscript. The authors thank Charlotte Simmonds for helpful comments which improved the manuscript. The authors gratefully acknowledge the Gauss Centre for Supercomputing e.V. (\url{www.gauss-centre.eu}) for funding this project by providing computing time on the GCS Supercomputer SuperMUC-NG at Leibniz Supercomputing Centre (\url{www.lrz.de}), under project pn29we. WM thanks the Science and Technology Facilities Council (STFC) Center for Doctoral Training (CDT) in Data Intensive Science at the University of Cambridge (STFC grant number 2742968) for a PhD studentship. WM and ST acknowledge support by the Royal Society Research Grant G125142. RK acknowledges support of the Natural Sciences and Engineering Research Council of Canada (NSERC) through a Discovery Grant and a Discovery Launch Supplement (funding reference numbers RGPIN-2024-06222 and DGECR-2024-00144) and York University's Global Research Excellence Initiative. EG is grateful to the Canon Foundation Europe and the Osaka University for their support through the Canon Fellowship. Support for OZ was provided by Harvard University through the Institute for Theory and Computation Fellowship. XS acknowledges the support from the National Aeronautics and Space Administration (NASA) grant JWST-AR-04814.

\section*{Data Availability}

All simulation data, including snapshots, group, and subhalo catalogs and merger trees will be made publicly available in the near future. Data will be distributed via \url{www.thesan-project.com}. Before the public data release, data underlying this article will be shared on reasonable request to the corresponding author(s).



\bibliographystyle{mnras}
\bibliography{main} 




\appendix


\bsp	
\label{lastpage}
\end{document}